# The Network Structure of Economic Output

Ricardo Hausmann[1,2] and César A. Hidalgo[2,3]

[1] Harvard Kennedy School, Harvard University
[2] Center for International Development, Harvard University
[3] Macro Connections, The MIT Media Lab, Massachusetts Institute of Technology

## Abstract:

Much of the analysis of economic growth has focused on the study of aggregate output. Here, we deviate from this tradition and look instead at the structure of output embodied in the network connecting countries to the products that they export. We characterize this network using four structural features: the negative relationship between the diversification of a country and the average ubiquity of its exports, and the non-normal distributions for product ubiquity, country diversification and product co-export. We model the structure of the network by assuming that products require a large number of non-tradable inputs, or capabilities, and that countries differ in the completeness of the set of capabilities they have. We solve the model assuming that the probability that a country has a capability and that a product requires a capability are constant and calibrate it to the data to find that it accounts well for all of the network features except for the heterogeneity in the distribution of country diversification. In the light of the model, this is evidence of a large heterogeneity in the distribution of capabilities across countries. Finally, we show that the model implies that the increase in diversification that is expected from the accumulation of a small number of capabilities is small for countries that have a few of them and large for those with many. This implies that the forces that help drive divergence in product diversity increase with the complexity of the global economy when capabilities travel poorly.

**JEL Codes:** O11, O14, O33, O57, F43, F47
**Keywords:** Diversification, Ubiquity, Capabilities, Poverty Trap, Economic Complexity, Structural Transformation, The Product Space, Networks.
**Acknowledgments:** We would like to acknowledge comments from Pol Antràs, Dany Bahar, Elhanan Helpman, Robert Lawrence, Raja Kali, Lant Pritchett, Roberto Rigobon, Dani Rodrik, Adrian Wood, Muhammed Yildirim and Andrés Zahler. We also would like to acknowledge the comments from the audiences at the 2009 International Growth Week Seminar, from the International Growth Center, the CID Faculty Lunch, UNCTAD, the Eastern Economic Association 2011 and the PREM Seminar Series at The World Bank for their comments and questions.

# I. Introduction

Making sense of aggregate growth with aggregate models has been difficult. Ever since Robert Solow's seminal papers (1956, 1957) it has been clear that growth cannot be explained as the consequence of the accumulation of aggregate factors such as physical or human capital, but that something else is at stake, something that affects the productivity of economic activities. As argued in Lucas (1988) "*a successful theory of development (or anything else) has to involve more than aggregative modeling.*" In fact, Lucas later argued that *"A growth miracle sustained for a period of decades thus must involve the continual introduction of new goods, not merely continual learning on a fixed set of goods."*

If this is so, then countries that have grown more in the past should have also introduced more products, so that today they should be more diversified. For example, Cameroun, Chile and the Netherlands had essentially the same population in the year 2000, about 15 million. However, when we look at their GDP per capita in dollars at market rates we find that they were respectively $635, $4917 and $24,180. In the same year, out of the 772 products in the SITC-4 Rev 2 classification (Robert C. Feenstra et al., 2005), Cameroun, Chile and the Netherlands had non-zero exports in 91(12%), 487(63%) and 745 (96%) items[1]. This suggests that, after controlling for population, rich countries appear to have introduced more goods than poor countries.

Developing models with many goods and goods that differ in some fundamental way, however, has not been easy. The basic workhorse on which many growth models have been based is the Dixit-Stiglitz production function (A. K. Dixit and J. E. Stiglitz, 1977), e.g. Grossman and Helpman (1991) or Aghion and Howitt (1992). The DS production function allows for closed form solutions and it is helpful in developing our intuitions. Yet, as pointed out by Krugman (2009): *"There is no good reason to believe that the assumptions of the Dixit-Stiglitz model – a continuum of goods that enter symmetrically into demand, with the same cost functions, and with the elasticity of substitution between any two goods both constant and the same for any pair you choose – are remotely true in reality."*

In fact, assuming that goods are fundamentally identical is problematic. Products seem to differ widely in the number of countries that successfully export them, suggesting that they are not equally easy to develop and that they differ in important ways that are not captured in the symmetries assumed by DS. The data shows that both, bookbinding machinery and polyurethanes, are products that tend to be exported successfully by fewer than 10 countries. Undergarments knitted of cotton, or wood of non-coniferous species, sawn, planed or tongued, however, are products exported successfully by most countries in the world. Moreover, the products exported by Cameroun, Chile and the Netherlands also differ in the number of other countries that on average have non-zero exports on them, respectively 87, 61 and 41 other countries. This suggests that there is something intrinsically different about the set of products that countries make that goes beyond their number.

Growth empirics have seen the introduction of many variables on the right hand side of growth equations, such as those related to human capital, institutions, geography or others. Yet, the left hand side has remained pretty much the same, in spite of Lucas's insistence regarding aggregation. Typically this includes some measure of growth in GDP per capita. In this paper we aim to accomplish two tasks. First, we aim to describe and compare economies in a manner that eschews aggregation. We do this by using network science, instead of aggregative models, and identify new stylized facts about both, the economic features of the world and that of individual countries. And second, we develop a very basic non-aggregative model of the network connecting countries to the products that they make and/or export and calibrate it to the data in order to account for the uncovered facts.

[1] In the Harmonized System 6 digit classification for 2009, out of 5111 items, Cameroun had non-zero exports in 1009 (19.7%), Chile in 2910 (56.9%) and the Netherlands in 5002 (97.9%).

Our step towards non-aggregative empirics and theory is made possible by the information contained in disaggregated trade data, which we interpret as a bi-partite network connecting countries to the products that they make or export. We use trade data, given its richness in terms of world coverage and product detail, yet the facts we focus on are unrelated to international trade per se, or to the production of goods rather than services. We support this by showing that a similar structure is found in the network connecting Chilean municipalities to the industries that they host, including both services and non-tradables.

Figures 1a and 1b show the matrix connecting countries to the products that they export. Here, the entries are the revealed comparative advantage of country *c* in product *p*. These matrices have a strong triangular shape, which is somewhat surprising since it implies that poorly diversified countries make products that most other countries make, while highly diversified countries make those products plus the products that few other countries make.

The matrix is triangular rather than block diagonal, as might have been expected by a simple interpretation of theories of trade based on factor proportions, suggesting a fundamental fact about the world: that the diversification of countries is inversely related to the ubiquity of the products that they make. We will show in this paper that this is a robust fact.

As shown in Hidalgo and Hausmann (2009), the information contained in the bi-partite network of products and countries correlates well with aggregate levels of per capita GDP, while the error terms of the relationship are predictive of future growth[2]. Why would the network contain information relevant for income and growth?

An answer to this puzzle is to assume that products are combinations of potentially many non-tradable inputs, which we call capabilities, and that countries make all the products for which they have the requisite capabilities. Products differ in the variety of capabilities they require and countries differ in the variety of capabilities they have. In formal terms, this means that the country-product network can be taken to be the result of the product of two other matrixes or networks: a country-capability network that expresses the endowment of capabilities of each country and a capability-product matrix that contains the technological requirement of products. Intuitively, countries with more capabilities will have what it takes to make more products, i.e. they will be more diverse. Products that require more capabilities will be accessible to fewer countries, i.e. will be less ubiquitous. Countries with more capabilities will be able to make products that require more capabilities, but these are less ubiquitous. Hence, more complex countries will be both more diversified and would make on average less ubiquitous products.

[2] This result is similar in style to Hausmann, Hwang and Rodrik (2007) who calculate a measure of export sophistication as the weighted GDP per capita of the countries that export a similar basket as the country in question and also show that it is predictive of GDP per capita and the error terms correlate with future growth. However, this paper uses information on GDP per capita to calculate export sophistication, while Hidalgo and Hausmann (2009) use only the links between countries and products. This implies that the result depends on the structure of the network and not on the GDP measures.

We formalize this way of thinking by introducing a baseline model in which we assume that the distribution of the country-capability and product-capability matrixes are random, and posit an operator that assumes that countries will make all products for which they have the requisite capabilities. We calibrate the model with three parameters: the probability that a country has a capability, the probability that a product requires a capability, and the number of capabilities in the world. We show how this model fits the observed stylized facts.

Finally, we show that, under very general conditions, the model predicts a non-linear relationship between the number of capabilities that a country has and the number of products that it makes. Countries with few capabilities will have a lower probability of finding uses for any additional capability than countries with many capabilities as the number of potential combinations increases as a power of the number of capabilities available in a country. Hence, countries with few (many) capabilities will face low (high) incentives to the accumulation of additional capabilities. We call this the *quiescence trap.* Moreover, we find that the depth of the quiescence trap increases with the number of capabilities that exist in the world and with the fraction of capabilities that the average product requires. The calibration of the model to the empirical data suggests that our world is one in which the quiescence trap is strong, a fact which may help explain the divergence of incomes over the past two centuries, as small differences in initial capability endowments would be amplified over time.

**Relation to Literature on Trade Theory**

The fact that more diversified countries tend to export products that are on average less ubiquitous is not directly implied by existing trade models, which would need some additional ancillary hypotheses to account for these facts. Classical trade theory, whether of the Ricardian or the Heckscher-Ohlin type, tries to explain why countries specialize in different products. As such, these theories take production as given and attempt to explain which countries will find it advantageous to specialize in a particular set of goods. These theories, however, make no predictions about the number of products made by a country and about the number of countries that make a product. In other words, these theories do not make detailed predictions about the structure of the network connecting countries to the products they make or export, or even regarding the diversification of countries, the ubiquity of products, and the relationship between these two dimensions.

New trade theory, on the other hand, (E. Helpman and P.R. Krugmann, 1985, P. R. Krugman, 1979) was developed to account for intra-industry trade. At the basis of that explanation is the assumption that there are scale economies in product development and that products are not homogeneous but differentiated. Because these varieties are imperfect substitutes, firms have some market power, but competition erodes their profits so that the monopoly profits they generate in production barely cover the fixed cost of product development. Larger countries have bigger markets in which to amortize the fixed costs of product development and thus would tend to be more diversified. Schott (2004) and Hummels and Klenow (2005) provide evidence of this effect.

New trade theory, however, makes no predictions about *which* products will be developed in each country. This is because, as argued above, the theory uses the Dixit-Stiglitz model (A. K. Dixit and J. E. Stiglitz, 1977) which posits a continuum of goods and makes strong assumptions about the symmetry of all goods in order to allow for simple closed-form solutions that are analytically tractable. This eliminates any intrinsic characteristic of the goods considered.

There are several elements about the world that get abstracted from view in the DS world. First, the cost of product development is independent of any characteristic of the product, since they are all the same. Second, the cost is also independent of the relationship between a particular product and the previous productive history of the country. For instance, the cost of developing a regional jet aircraft is the same whether the firm or country has previously developed a transcontinental aircraft and a combustion engine or whether it produces only raw cocoa and coffee. To make this class of models compatible with our stylized facts, one would have to abandon the idea of the continuum of products and look instead at the varying density of products. In addition, the cost of developing a new variety would not be a constant but instead would depend on the nature of the products already present. Building on these ideas Kali et al. (2010) have recently advanced a model in this direction.

Similarly, the Dixit-Stiglitz production function has found its way into theories of growth, where productivity is related to the number of intermediate inputs countries have available for production, with the assumption that the greater the number of intermediate inputs, the higher the productivity with which the economy can operate (A. Rodriguez-Clare, 2007) (D. Acemoglu et al., 2007). Again, the DS production function assumes that the cost of developing new intermediate inputs is independent of the quantity and nature of the intermediate inputs that are already available, making the growth process independent of the specific structure of production. Also the link between the number of intermediate inputs and productivity is assumed, not explained by these models.

The Melitz trade model (M. J. Melitz, 2003), on the other hand, assumed firm heterogeneity to explain trade data at the firm level, but suffers from similar limitations in terms of the assumptions about technology that prevent it from explaining a systematic relationship between the diversification of countries and the ubiquity of its products. The fact that the $\boldsymbol{M}_{cp}$ matrix is triangular and not block diagonal suggests that firm heterogeneity across borders is large, explaining why the same products are exported by countries with radically different relative prices.

### Relation to Literature on Economic Growth and Development

When it comes to growth theory, our approach is related to the recombinant growth model introduced by Weitzman (1998), which is in itself highly related to the grammar model introduced by Kauffman (1993). In both, Weitzman's and Kauffman's models, the development of new varieties emerges as combinations of previous varieties. Both knowledge of chemistry and optics are required to create photography.

In the formalism that we introduce later, this can be interpreted loosely as an increase in the total number of capabilities that exist in the world. Our model differs from that of Weitzman and Kauffman, however, in various dimensions. First, we do not model the historical number of potential varieties that exist in a world, but rather the number of feasible varieties that countries can produce given a limited capability endowment. Second, we use our model to explain differences in the diversity of countries, the ubiquity of products, the connection between these two variables, and the probability that a pair of products would be co-exported. Weitzman uses his model to explain the lack of acceleration implied by endogenous growth theory, as an information problem, whereas Kauffman uses his grammar model to explain the historical increase in product diversity. Neither of them, however, outline the implications of their models for the differences in diversity and ubiquity observed in the world. Finally, since the models presented by both Kauffman and Weitzman do not consider connections between countries and products, they do not make predictions about either the structure of the network connecting countries to the products they export, or its projection into the space of products.

Due to the nature of our observables, our approach deviates from the class of models started by Robert Solow (1956) half a century ago and continued in the endogenous growth theory, including Romer (1986) or Kremer's O-ring model (1993). The O-ring model assumes that products differ in the *number* of complementary steps that they require where each step is otherwise identical. In this model, countries with greater ability to perform any step successfully will find it more advantageous to specialize in products that require many steps. Yet, they will be unable to compete with less able countries that specialize in products requiring fewer steps, since wage differentials would make the production of these goods in the most able countries too costly. This model would not predict that high ability countries would be more diversified per se and thus cannot account for the basic stylized fact uncovered in this paper. Indeed, the O-ring model predicts that the matrix connecting countries to products would take a block diagonal rather than a triangular shape.

Our empirical approach is based on the idea that products require the local availability of a potentially large set of non-tradable factors of production, which we call capabilities. We assume that information about “which country makes what” carries information about which country has which capabilities. Hidalgo and Hausmann (2009) showed that it is possible to count the relative number of capabilities in a country, without making any assumptions regarding their nature, by creating measures that incorporate information that combine the diversification of countries and the ubiquity of products. These measures of input, or capability diversity, were shown to correlate strongly with GDP per capita and have residuals that predict future economic growth, suggesting that countries tend to approach a level of income which is determined by their capability endowment.

In this framework, we look at the process through which diversification increases. Countries diversify by accumulating new capabilities and using these, in combination with others, for the production of new products. Hidalgo and Hausmann (2009) show that countries with many capabilities (countries that are highly diversified and make non-ubiquitous products that other diversified countries tend to make), are more likely to add products that require many capabilities to their export basket (i.e. that have low ubiquity and are exported by highly diversified countries). The reverse holds true for countries with few capabilities. This suggests that the mix of products made by a country increases gradually through the addition of capabilities, and that this gradual development leaves fingerprints in the structure of the network connecting countries to products.

Indeed, as shown in Hidalgo et al. (2007), the likelihood that a country develops a particular product depends on how "near" is that product in the "product space" to the products that the country is already able to successfully export. Here proximity is related to the probability that those two goods are co-exported in other countries. The product space, however, is highly heterogeneous. Its sparse sections and dense patches imply that the world does not exhibit the symmetries assumed by DS and that not all countries are similarly located. Countries that are better positioned in the product space, in the sense of having more nearby products, tend to have better opportunities to diversify and tend to outgrow countries that produce products that are less connected (R. Hausmann and B. Klinger, 2006). The shape of the product space can also be used to explain the lack of convergence in the world economy (Hidalgo et al. 2007) since there are distances between products in the space that are larger than the distances countries are empirically able to traverse. Also, the presence of nearby products is associated with the resilience of economies to external shocks. Hausmann, Rodriguez, and Wagner (2006) find that countries that are more disconnected in the Product Space tend to suffer more frequent, longer and deeper recessions than countries that are more centrally positioned in this network.

**Relation to Literature on Measures of Diversification or Concentration**

The incorporation of information on product ubiquity differentiates our measure from other measures of diversification, such as the Herfindahl-Hirschman (1964) index or entropy (L. Jost, 2006, P. P. Saviotti and K. Frenken, 2008). Both the HH-index and entropy do not include any information on products, making their measures of diversification identical for any two baskets of goods that have the same distribution of shares. In other words, both the HH-index and entropy cannot distinguish between countries exporting 10% bananas and 90% mangos, 90% bananas 10% mangos, or 10% motorcycles and 90% aircraft engines.

The remainder of the paper is organized in the following way. In the next section we provide details on the data used. Section II introduces our network analysis of the data and established the stylized facts observable in the data. Section III introduces the model and solves it for the particular case in which the probability that a country has a capability and the probability that a capability is required to make a product are both constant and equal across countries and products. Section IV calibrates this model to the data. Section V draws implications of the model and Section VI concludes.

## II. Methods, Data and Stylized Facts

We use trade data to connect countries to the products that they export. To show that the results presented below are not driven by any particular form of encoding products, we use two different trade classifications systems. The Feenstra et al. dataset (2005), which codes products using the SICT4 rev2 (1006 products) classification and the Base pour l'Analyse du Commerce International (BACI) dataset from the Centre d'Études Prospectives et d'Informations Internationales (CEPII), which contains data for 232 countries and 5,109 product categories classified using the Harmonized System at the 6-digit level (G. Gaulier and S. Zignano, 2009).

To make countries and products more readily comparable, we control for variations in the size of countries and of product markets by calculating the Revealed Comparative Advantage (RCA) that each country has in each product. For this we use Balassa's (1964) definition of RCA as the ratio between the export share of product $p$ in country $c$ and the share of product $p$ in the world market. Formally RCA is defined as:

$$RCA_{cp} = \frac{X_{cp}}{\sum_p X_{cp}} \Big/ \frac{\sum_c X_{cp}}{\sum_{c,p} X_{cp}}, \qquad (1)$$

where $X_{cp}$ represents the dollar exports of country $c$ in product $p$.

To show that the results we uncover are not the consequence of differences between production and exports, or between goods and services, we use a datasets that connects Chile's 347 municipalities to firms in 700 different industrial categories that include all sectors. The Chilean dataset, however, will not be studied in detail in this paper.

## Four Stylized Facts

We study the structure of the bipartite network connecting locations and industries (countries and products) by defining the adjacency matrix $M_{cp}$ as:

$M_{cp}$=1 if country $c$ exports product $p$ with an RCA above a certain threshold. Going forward, we denote an RCA thresholds as $R^*$.
$M_{cp}$=0 otherwise.

We define the diversification of country $c$ as the sum of $M_{cp}$ over all products

$$k_{c,0} = \sum_p M_{cp} \qquad (2)$$

and the ubiquity of product $p$ as the sum of $M_{cp}$ over all countries

$$k_{p,0} = \sum_c M_{cp} \qquad (3)$$

Finally, we define the average ubiquity of the products exported by country $c$ as:

$$k_{c,1} = \frac{1}{k_{c,0}} \sum_p M_{cp} k_{p,0}, \qquad (4)$$

and the average diversification of a product's exporters as

$$k_{p,1} = \frac{1}{k_{p,0}} \sum_c M_{cp} k_{c,0}. \qquad (5)$$

**Stylized Fact 1: Diversification and Ubiquity are inversely related.**

As argued above, a conspicuous fact of the structure of the network connecting countries to the products that they make or export is that poorly diversified countries export products that are, on average, exported by many other countries, whereas highly diversified countries make products which are made, on average, by fewer other countries. Figures 1a and 1b present matrices showing the RCA that countries have on products. Rows represented countries and are sorted according to their *diversification*, whereas columns represented products and sorted by their *ubiquity*. These matrices exhibit a triangular shape.[3] Figure 2 shows the same pattern using Chilean production data, which includes information for the whole economy, not just exported goods[4].

Formally, we quantify this stylized fact, and demonstrate that it is not implied trivially by the heterogeneity in the distribution of country diversification and product ubiquity, by introducing two diagrams: the $k_{c,0}$-$k_{c,1}$ diagram (country diversification–average ubiquity of its products) and the $k_{p,0}$-$k_{c,p}$ diagram (product ubiquity-average diversification of its exporters).

Figure 1c and 1d show the $k_{c,0}$-$k_{c,1}$ (diversification–average ubiquity) diagrams corresponding to the RCA matrices shown in Figure 1 for $R^*$=1. In all cases, we observe that the average ubiquity of a country's exports tends to decrease with that country's level of diversification. Countries such as the US and Germany are in the lower right-hand side of the graph and countries in the upper left are poor. A similar pattern occurs in the data for Chile. Santiago and its other municipalities are in the lower right-hand side and mostly rural remote and poor municipalities are in the upper left-hand side (Figure 2).

Because of the symmetry between countries and products that is inherent in $M_{cp}$, we define an equivalent diagram for products. In the case of products the $k_{p,0}$-$k_{p,1}$ diagram shows that the average diversification of the countries' exporting a product falls, on average, as a function of the ubiquity of that product (Figure 1 e and f).

We test the statistical significance of these patterns by introducing four null models (Fig 1g). Since these diagrams summarize structural properties of bipartite networks, their significance can be assessed only by comparing them to bipartite networks with equivalent structural properties (S. Maslov and K. Sneppen, 2002). **Null Model 1** is a random network with the same number of links, that is, with the same average ubiquity and diversification as $\boldsymbol{M_{cp}}$. **Null Model 2** is a randomized network in which the values inside each column of $\boldsymbol{M_{cp}}$ have been shuffled and represent a network in which the diversification of each country matches exactly that observed in the data, yet its exports have been randomly reassigned such that the average ubiquity of the system is conserved. **Null Model 3** is a randomized network in which the values in each row of $\boldsymbol{M_{cp}}$ have been shuffled and represent a network in which the ubiquity of each product matches exactly the one observed in the original data, but the producers of those products have been randomly assigned. The average diversification of Null Model 3 matches that of the original data. **Null Model 4** is a randomized network constructed by permuting the entries of $\boldsymbol{M_{cp}}$ such that the ubiquity of products and diversification of countries remains unchanged. Null Model 4 is the most stringent of the four null models, as it preserves exactly the diversification of each country ($k_{c,0}$) and the ubiquity of each product ($k_{p,0}$). Because of its stringency, however, Null Model 4 does not randomize countries producing or exporting a substantial fraction of all products and products that are being produced or exported by a large fraction of countries.

---

[3] To order these matrices we calculate diversification using (2) as the number of products that they export with an RCA above a certain threshold (taken as RCA≥0.5 in this example), and product *ubiquity* using (3) as the number of countries exporting a product with an RCA above a certain threshold (also taken to be RCA≥0.5 in this example). This is done only for the purpose of ordering the matrix. The actual values of RCA are color-coded.

[4] This relationship is related to the number-average-size rule described by T Mori et al. **Mori, T.; K. Nishikimi and T. E. Smith.** 2008. "The Number-Average Size Rule: A New Empirical Relationship between Industrial Location and City Size." *Journal of Regional Science*, 48(1), 165-211. Yet, the number-average-size rule relates the number and the average (population) size of metro areas in which a given industry is present.

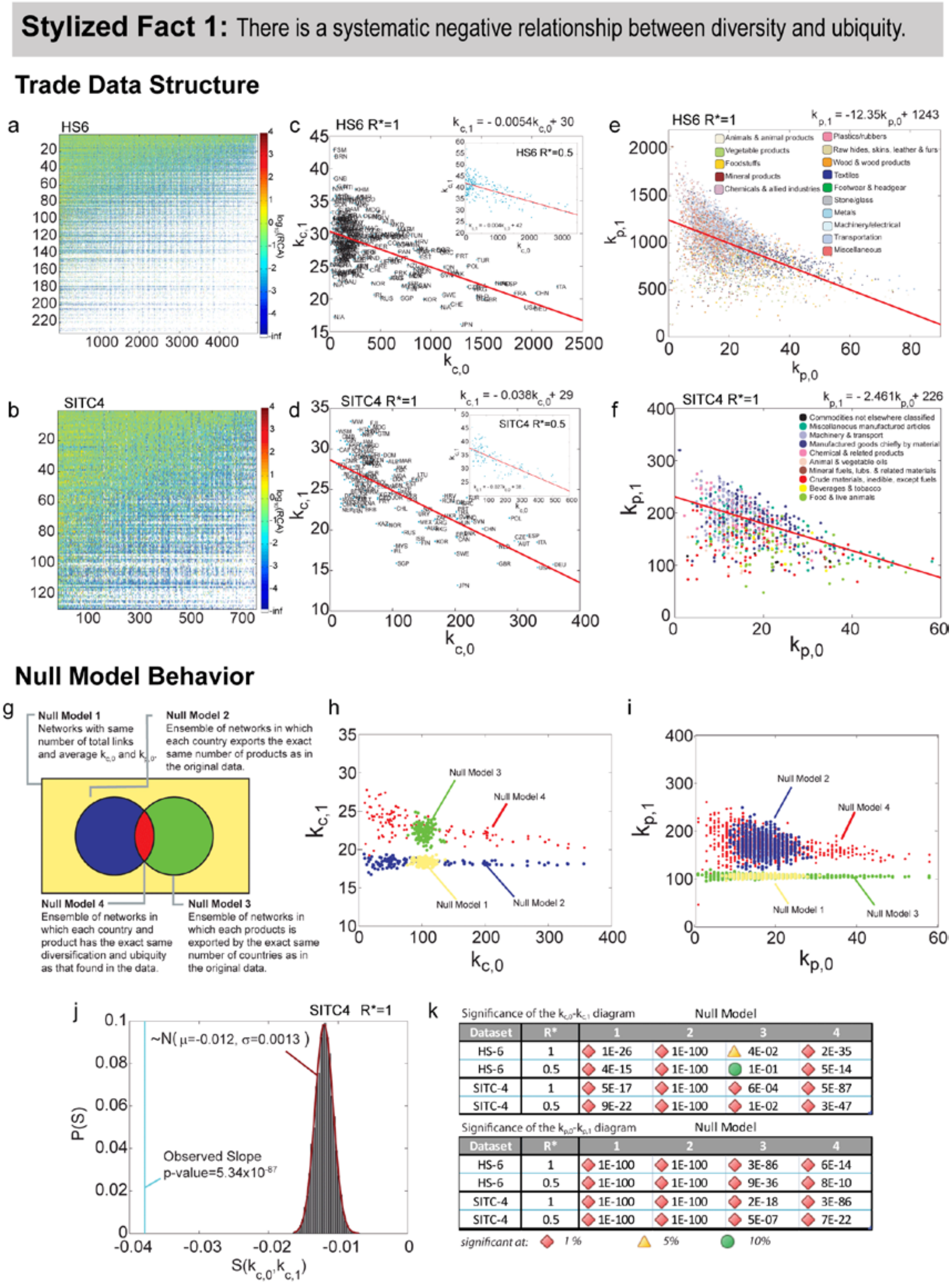

Significance of the $k_{c,0}$-$k_{c,1}$ diagram — Null Model

| Dataset | R* | 1 | 2 | 3 | 4 |
|---|---|---|---|---|---|
| HS-6 | 1 | 1E-26 | 1E-100 | 4E-02 | 2E-35 |
| HS-6 | 0.5 | 4E-15 | 1E-100 | 1E-01 | 5E-14 |
| SITC-4 | 1 | 5E-17 | 1E-100 | 6E-04 | 5E-87 |
| SITC-4 | 0.5 | 9E-22 | 1E-100 | 1E-02 | 3E-47 |

Significance of the $k_{p,0}$-$k_{p,1}$ diagram — Null Model

| Dataset | R* | 1 | 2 | 3 | 4 |
|---|---|---|---|---|---|
| HS-6 | 1 | 1E-100 | 1E-100 | 3E-86 | 6E-14 |
| HS-6 | 0.5 | 1E-100 | 1E-100 | 9E-36 | 8E-10 |
| SITC-4 | 1 | 1E-100 | 1E-100 | 2E-18 | 3E-86 |
| SITC-4 | 0.5 | 1E-100 | 1E-100 | 5E-07 | 7E-22 |

*significant at:* ◆ 1% ▲ 5% ● 10%

**Figure 1** Stylized Fact 1: Diversification and ubiquity are negatively related. **a**. Adjacency matrix of the HS6 Country Product Network (year 2005) sorted by the diversification of countries and the ubiquity of their exports calculated using $R^*$=0.5. The color in each entry of these matrices represents the logarithm, in base 10, of the RCA that countries (rows) have on products (columns). The mapping between colors and values can be read from the colorbar **b.** Same as a but using the SITC4 dataset for the year 2000. **c** The $k_{c,0}$-$k_{c,1}$ diagram for the HS6 dataset (year 2005) calculated for R*=1 and R*=0.5. **d** Same as c but for the SITC4 dataset for the year 2000. **e** The $k_{p,0}$-$k_{p,1}$ diagram for the HS6 dataset (year 2005) calculated for R*=1 and R*=0.5. **f** Same as **e** but for the SITC4 dataset for the year 2000. **g**. Explanation of the null models used to test the significance of these diagrams. **h**. The $k_{c,0}$-$k_{c,1}$ diagram and the **i** $k_{p,0}$-$k_{p,1}$ diagram for each of the null models. **j** Interpolation method used to compare the slopes of the null models with that observed in the data. **k** Table summarizing the statistical significance of the diagrams in **c-f** estimated using 1000 implementations of each null model.

We use the four null models described above to estimate a *p*-value for the probability of observing a slope of a certain magnitude in each of these diagrams. Figure 1j illustrates how this procedure was done and summarizes the *p*-values obtained for the three datasets and two RCA thresholds (Fig 1k). The method consists of creating 1000 different instances of the null model, calculating the slopes for each one of them ($S(k_{c,0},k_{c,1})$), and fitting a normal curve to the distribution of slopes obtained from the ensemble of null models[5]. From this fit, it is possible to estimate the probability of observing the slope characterizing each data set given the null model constraints. This test demonstrates that the sharp negative slopes observed in all of the datasets emerges not from the heterogeneity of the distributions of diversification and ubiquity, but rather as a consequence of a non-trivial pattern of connections between countries and products.

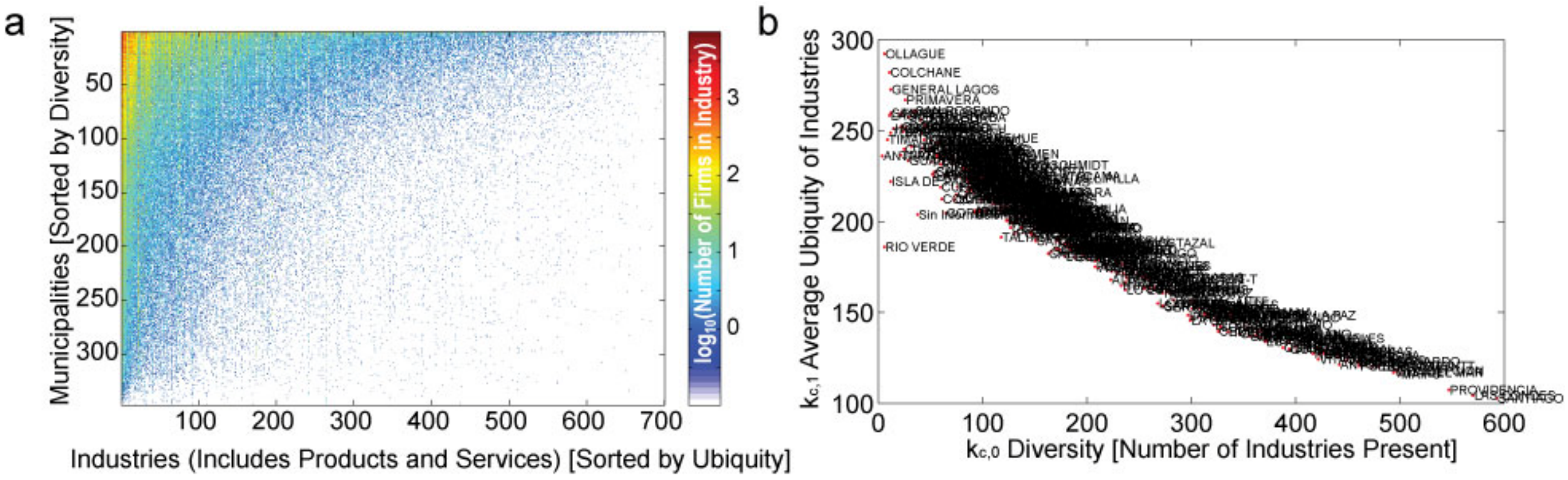

**Figure 2** Location-Industry matrix for Chile. **a** Location-Industry matrix, including all firms that are registered in the tax authority as paying taxes in a given municipality. Rows and columns are sorted as indicated in the figure and white indicates the complete absence of that industry in that municipality **b** Diversity-Average Ubiquity diagram for Chilean municipalities.

**Stylized Fact 2: Ubiquities are not normally distributed and their distribution is better approximated by a log-normal or Weibull distribution**

A stylized fact about the structure of the network connecting countries to the products that they export is that the distribution of ubiquities (number of countries that make a product), is not normal and it is better fitted by a log-normal or Weibull distribution. Figure 3 illustrates this using both datasets and two thresholds.

[5] The normal fit was included as an extrapolation method since most times the observed slope lay outside of the distribution defined by the 1000 null models and the normal distribution represented a good fit for the distribution of slopes emerging from the ensemble of null models.

MATLAB's distribution fitting tool was used to fit the ubiquity distribution to a log-normal distribution, a normal distribution and a two parameter Weibull distribution. We find that for all cases the normal distribution was the least likely fit and that for R*=1 the log-normal distribution was the most accurate approximation to the data. For R*=0.5, however, the Weibull distribution represents the best fit, with the log-normal as a close second. We take the departure of normality as relevant since normality is a standard null assumption and that the departure of the system from normality is a fact that theories need to account for.

**Stylized Fact 2:** The distribution of ubiquities is not normal, and it is better approximated by a log-normal or Weibull distributions

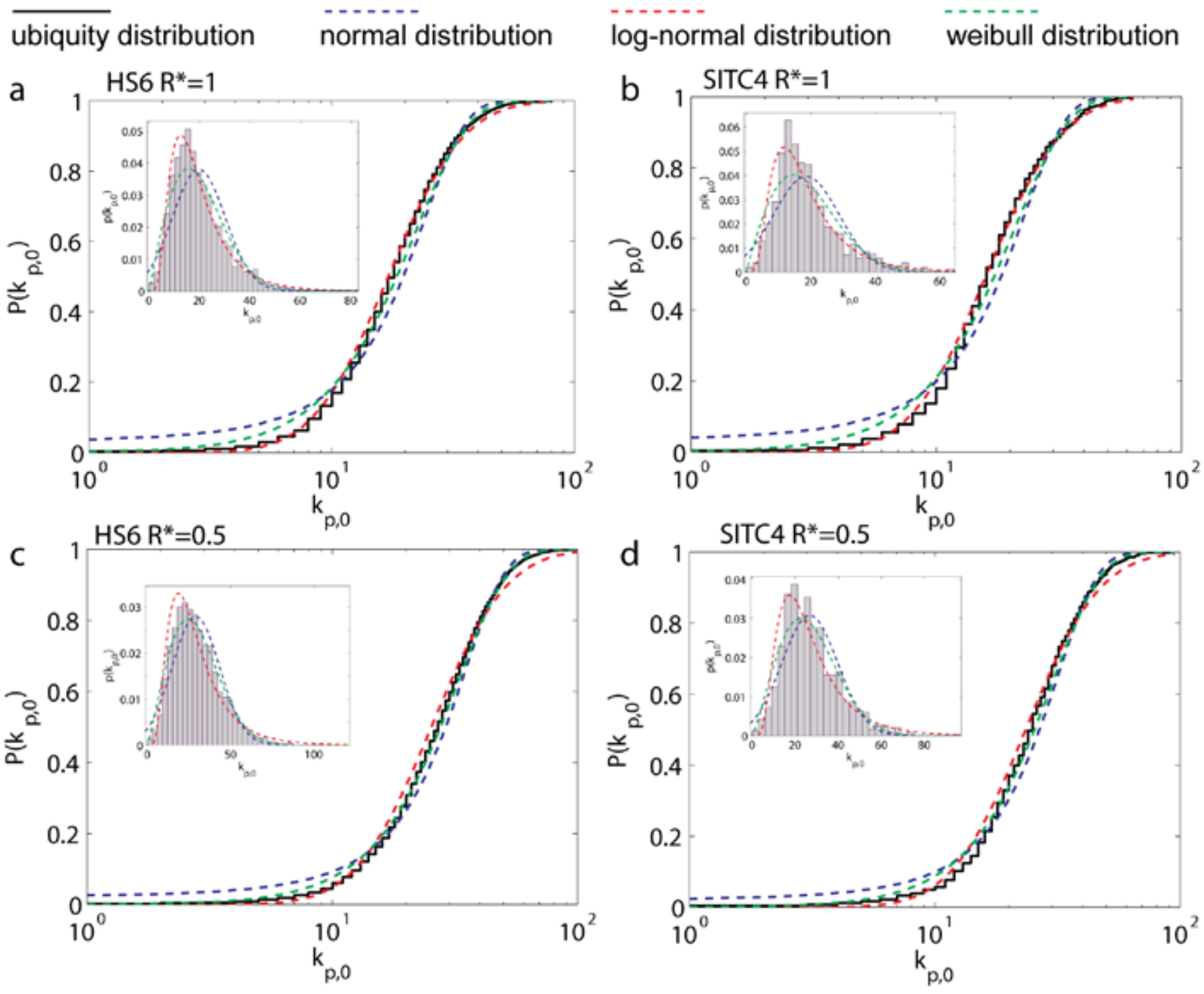

**Figure 3** The distribution of ubiquities. **a**-**d** Show the cumulative distribution function and the density distribution function for the ubiquity distribution obtained from the data. Dashed lines show the fits of normal, log-normal and Weibull distributions (see figure key for details). **a** HS6 R*1 **b** SITC4 R*=1 **c** HS6 R*=0.5 **d** SITC4 R*=0.5

### Stylized Facts 3: Diversifications are not normally distributed and their distribution is better approximated by a log-normal or Weibull distribution

The distribution of diversification (the number of products made or exported by a country), is not normal and it is better fitted by a log-normal or Weibull distribution. Figure 4 illustrates this using both datasets and two thresholds. The figures also show clearly that the distribution of diversification is much more heterogeneous than the distribution of ubiquities.

We use MATLAB's distribution fitting tool to find the most likely fit using a log-normal distribution, a normal distribution and a two parameter Weibull distribution. We find that the normal distribution was always the least likely fit and that the log-normal distribution and the Weibull distribution fitted the data much better. Moreover, we found that the difference between the likelihoods of the fits between a log-normal and a Weibull distribution were not highly statistically significant, indicating that both distribution represent a good approximation to the data.

Finally, we note that the departure of normality in the case of diversification is considerable larger than the departure of normality observed in the case of product ubiquity. This indicates an asymmetry in the structure of the network connecting countries to the products that they export.

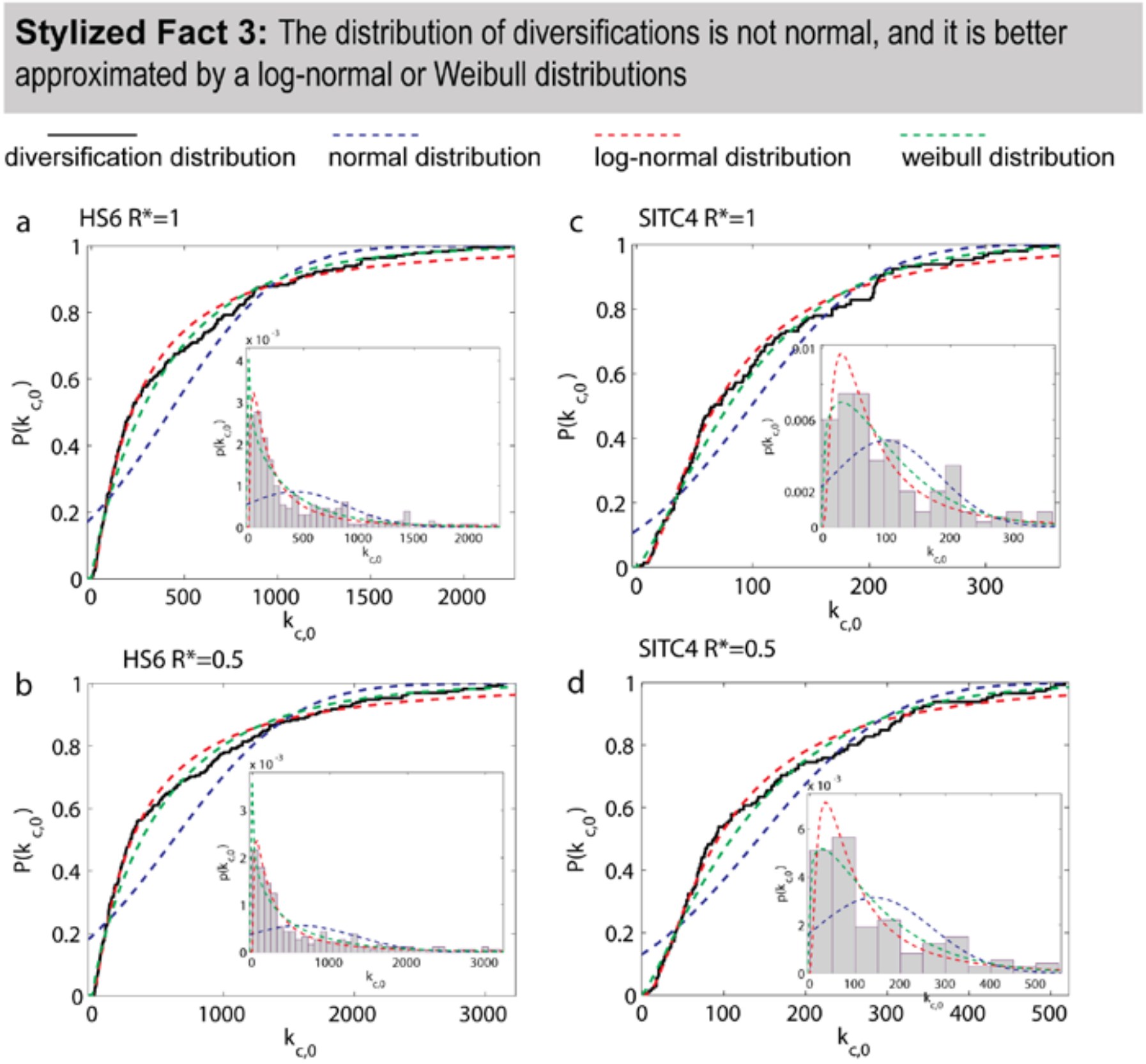

**Figure 4** The distribution of diversifications. **a**-**d** Show the cumulative distribution function and the density distribution function for the diversification distribution obtained from the data. Dashed lines show the fits of normal, log-normal and Weibull distributions (see figure key for details). **a** HS6 R*1 **b** SITC4 R*=1 **c** HS6 R*=0.5 **d** SITC4 R*=0.5

**Stylized Fact 4: The distribution of co-export proximities is not normal, and it is well approximated by a Weibull distribution**

The last stylized fact presented in this paper is the probability that a pair of products is co-exported. If two products require a similar set of capabilities, countries that are able to successfully export one of them are more likely to also be able to export the other. Hence, the information on the similarity of input requirements is contained in the pattern of co-exports of products. Hidalgo et al. (2007) have shown that the set of products that a country exports evolves over time following the structure of the network of product similarity, or product space.

We study the pattern of product co-exports and define as proximity the minimum of the conditional probability of exporting a product given the export of another good. A proximity value of 0.4, for instance, indicates that the probability that a country is exporting products $p$ and $p'$, given that it exports one or the other, is at least 40%. Using the matrix notation introduced above the proximity between products $p$ and $p'$ is defined as

$$\boldsymbol{\phi}_{pp'} = \frac{\sum_c M_{cp} M_{cp'}}{\max\,(k_{p,0,} k_{p',0,})} \tag{6}$$

Figure 5 shows the empirical distribution of co-exports for the two datasets presented above together with three fits: A normal distribution, a log-normal distribution and a two parameter Weibull distribution. After calculating the likelihood of each distribution for all of the three datasets, and the two thresholds, we find that the normal distribution is always the less likely fit and that the two parameter Weibull distribution is always the most likely fit, followed by the log-normal distribution.

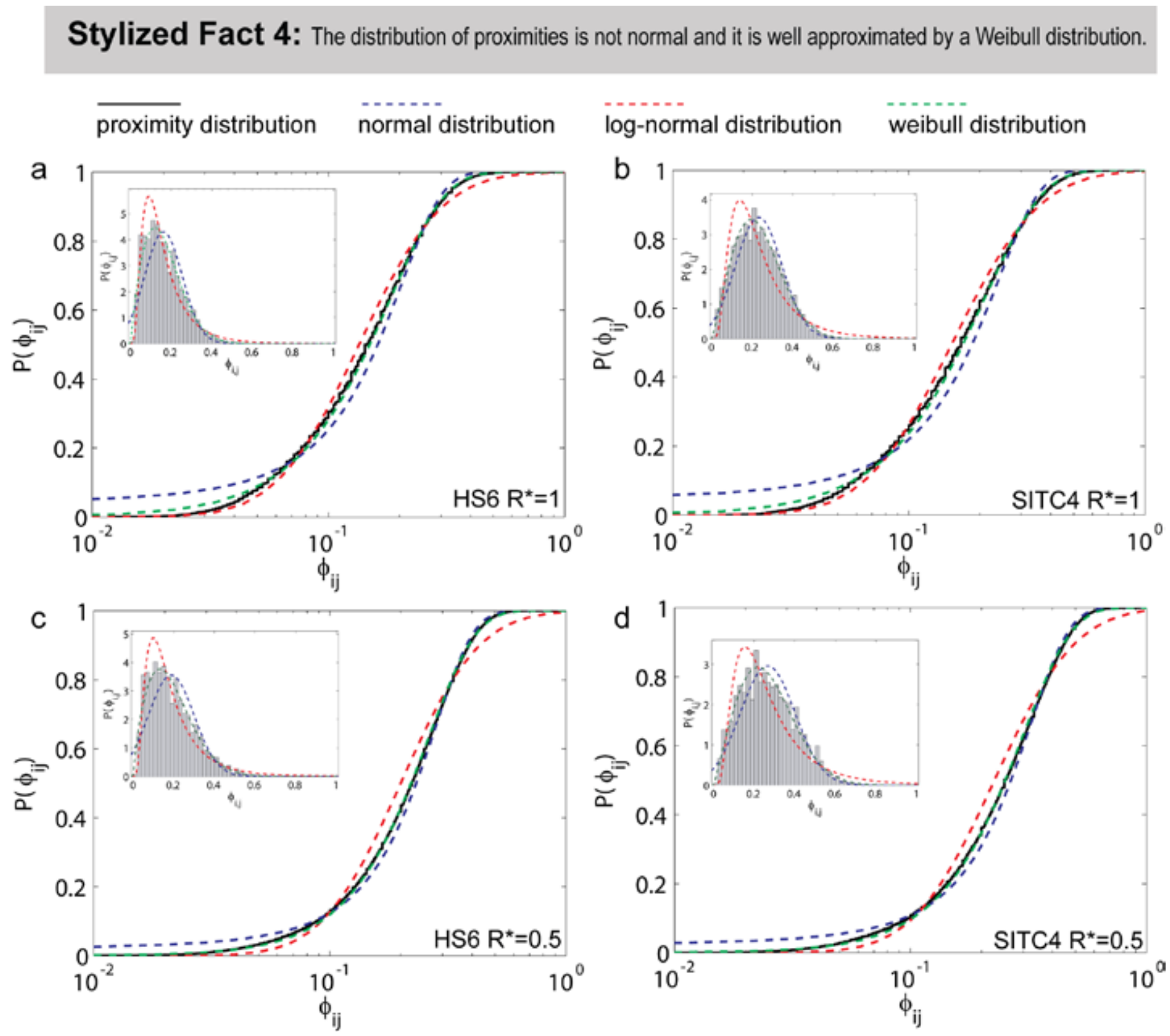

**Figure 5** The distribution of proximities. **a-d** Show the cumulative distribution function and the density distribution function for the proximity distribution obtained from the data. Dashed lines show the fits of normal, log-normal and Weibull distributions (see figure key for details). **a** HS6 R*1 **b** SITC4 R*=1 **c** HS6 R*=0.5 **d** SITC4 R*=0.5

## III. A Simple Modeling Framework

Here we introduce a simple modeling framework that can be used to understand and reproduce the global patterns of exports summarized in $\boldsymbol{M}_{cp}$. The model is based on the assumption that production requires the combination of a potentially large number of specific inputs, or capabilities, and that countries can only produce the goods for which they have all required capabilities. Mathematically, we describe a country as a vector, or list of adjacencies, which are equal to 1 if that country has that capability, and 0 otherwise. By the same token, products are described by the set of capabilities that they require, which can also be expressed using a vector in which 1's indicate the capabilities required to produce that product.

The world is represented using two matrices or networks: A country-capability matrix $\boldsymbol{C}_{ca}$, in which each row summarizes the capability endowment of country *c*; and a product-capability matrix $\boldsymbol{P}_{pa}$, in which each row summarizes the capability requirements of product *p*. The matrix connecting countries to the products that they make or export $\boldsymbol{M}_{cp}$ is assumed to be the result of a combination of these two matrices. Within this theoretical framework, we will refer to a model as a particular choice of a country-capability network, a product capability network and an operator taking both of these networks into a country-product network. Hence, the observable being modeled is the network connecting countries to products ($M_{cp}$) and the inputs of the model are the country capability matrix ($\boldsymbol{C}_{ca}$), the product capability matrix ($P_{pa}$) and the operator that is used to project these two networks into that connecting countries to products. Our use of the word model here differs from its traditional use in statistics (a functional form used for a regression) and in aggregative modeling (constrained optimization problem). Without loss of generality, we assume a world composed by $N_c$ countries, $N_p$ products, and $N_a$ capabilities.

In this interpretation, products require the combination of several inputs, some quite general, but others more specific to a smaller set of products. For instance, a shoe manufacturer and a circuit board company both need accountants and a cleaning crew, yet the shoe factory requires workers who are skilled in leather tanning and crusting, as well as leather cutting, sawing, and pasting. The circuit board manufacturing plant, on the other hand, does not need expert leather tanners or seamstresses, but requires people skilled in photo-engraving or PCB milling techniques, which have no use in the shoe factory. Each one of these requirements can be thought of the 1's and 0's which are specified in $\boldsymbol{P}_{pa}$. Yet, in general, we can think that these binary entries include specific infrastructure, regulations, norms, and other non-tradable activities, such as port or postal services, whose presence or absence can either facilitate or limit the production of these products. Indeed, the formalism we present next helps track the implications of assuming that countries and products differ in the set of capabilities they have or require without requiring any definition of what these capabilities are.

Moreover, we assume that each of these products, defined narrowly enough, cannot be produced in the absence of any of the inputs that need to be locally available. This defines $\boldsymbol{C}_{ca}$. For instance, "tanned leather" cannot be produced without leather tanners and "women shirts" cannot be produced without seamstresses. Hence, we consider that the production of "tanned leather" by a country suggests the existence of leather tanners in it. This assumption by no means implies that there are no possible substitutions between capabilities. This is because capabilities can be grouped together until a set of purely complementary capabilities is reached and no further substitutions are possible. We assume to be working in that renormalized limit. We also assume that inputs that can be easily imported do not pin down where production can take place, so we put the emphasis on non-tradable capabilities.

### The Binomial Model

In this paper we concentrate on the study of a specific form of the operator used to combine the country-capability matrix and the product-capability matrix into the country-product matrix. This is a Leontief-like production function in which the production of each of these products will take place if all the requisite inputs are present and will be equal to zero in absence of any of them. Alternatively, we could think of this operator as a discrete version of a Constant Elasticity of Substitution (CES) production function, but with many potential inputs. If a country lacks any of the inputs that go into a product, output will be zero.

Formally we denote this operation as:

$$\boldsymbol{M}_{cp} = \boldsymbol{C}_{ca} \odot \boldsymbol{P}_{pa}, \qquad (7)$$

where

$$\boldsymbol{C}_{ca} \odot \boldsymbol{P}_{pa} = 1 \text{ if } \sum_{a}^{N_a} \boldsymbol{C}_{ca}\boldsymbol{P}_{pa} = \sum_{a}^{N_a} \boldsymbol{P}_{pa} \text{ and 0 otherwise.} \qquad (8)$$

We refer to this particular form of the $\odot$ operator as the Leontief operator[6], because it resembles a Leontief production function, but in a binary form. This operator can also be thought of as the subset operator, since $\boldsymbol{M}_{cp}$=1 if the capabilities required by product *p* are a subset of the capabilities present in country *c*.

Hence, we do not assume any mechanisms that would force countries to specialize other than the availability of capabilities. If a country has the capabilities we assume it will make all the products that are feasible with these capabilities. This goes against the grain of what much of classical trade theory was about, but the triangular shape of the RCA matrixes suggest that there is little specialization, even at the level of 5100 products.

We consider the particular case in which $\boldsymbol{C}_{ca}$=1 with probability *r* and 0 with probability 1-*r*, and $\boldsymbol{P}_{pa}$=1 with probability *q* and 0 with probability 1-*q*.

Here we do not adopt any a priori definition of capabilities and therefore consider $\boldsymbol{C}_{ca}$ and $\boldsymbol{P}_{pa}$ as empirically unobservable quantities. Our goal is to illustrate how the structure of $\boldsymbol{M}_{cp}$ can be deduced from the structure of $\boldsymbol{C}_{ca}$ and $\boldsymbol{P}_{pa}$ and will compare the $\boldsymbol{M}_{cp}$ implied by our binomial implementation of this theory with the empirically observed one. For this we will compare the matrix with the four observables presented above in the next section.

To differentiate between the number of links connecting a country to a product we use the superscripts (*a*) for capabilities. Hence, we define,

$$Number\ of\ Capabilities\ present\ in\ c: k_{c,0}^{a} = \sum_{a=1}^{N_a} \boldsymbol{C}_{ca} \qquad (9)$$

---

[6] We note that it is mathematically possible to rewrite (8) using regular matrix multiplication, yet this requires redefining matrices and lead to a representation that is not as simple as the one presented here.

$$Number\ of\ capabilities\ required\ by\ p\colon k_{p,0}^{a} = \sum_{a=1}^{N_a} \boldsymbol{P}_{pa} \tag{10}$$

**Mean Field Estimates for The Binomial Model**

In this section we use mean field approximations to predict the functional forms that we expect to emerge from the model. First, we look for a function that relates the average level of diversification of a country to the number of capabilities it has. Namely we are searching for

$$\overline{k_{c,0}}(k_{c,0}^{a}). \tag{11}$$

where overhead bars are used to denote averages, or expected values.

We calculate the diversity of a country with $k_{c,0}^{a}$ capabilities by adding the number of products requiring a given number of capabilities times the probability that a country will export such good. Mathematically we represent this in a general form as:

$$\overline{k_{c,0}} = \sum_{x=0}^{N_a} \pi\left(c(k_{c,0}^{a}) \rightarrow p(k_{p,0}^{a} = x)\right) N_p(k_{p,0}^{a} = x), \tag{12}$$

where $\pi\left(c(k_{c,0}^{a}) \rightarrow p(k_{p,0}^{a} = x)\right)$ represents the probability that a country *c*, with $k_{c,0}^{a}$ capabilities exports a product *p* requiring $k_{p,0}^{a} = x$ capabilities. The number of products requiring *x* capabilities is represented by $N_p(k_{p,0}^{a} = x)$.

We calculate the expected diversification of country *c* ($\overline{k_{c,0}}$) by considering the number of capabilities that the country has as given and equal to: $k_{c,0}^{a}$. This implies that the realization of the random variable *r* for that country is equal to the number of capabilities it has over the number of capabilities that exist, or $(k_{c,0}^{a}/N_a)$. Additionally, from the Leontief operator, the probability that a country with $k_{c,0}^{a}$ capabilities exports a product requiring $k_{p,0}^{a} = x$ capabilities, is given by the probability that that country has all the capabilities required by that product. As in this model, the capabilities that a country has are independent random variables, the probability that a country has the *x* capabilities required to produce a product is given by:

$$\pi\left(c(k_{c,0}^{a}) \rightarrow p(k_{p,0}^{a} = x)\right) = \left(\frac{k_{c,0}^{a}}{N_a}\right)^{x}. \tag{13}$$

Now, since products require a capability with a probability *q*, the number of products requiring *x* capabilities is given by a binomial distribution (which is why we call this implementation the binomial model). Hence,

$$N(k_{p,0}^{a} = x) = N_p \binom{N_a}{x} q^{x}(1-q)^{N_a - x}. \tag{14}$$

Using (13) and (14), we can take (12) into the model specific form

$$\overline{k_{c,0}} = N_p \sum_{x=0}^{N_a} \left(\frac{k_{c,0}^{a}}{N_a}\right)^{x} \binom{N_a}{x} q^{x}(1-q)^{N_a - x}, \tag{15}$$

which can be simplified using the binomial theorem, or Newton's Binomial, to

$$\overline{k_{c,0}} = N_p \left( q \frac{k_{c,0}^a}{N_a} + 1 - q \right)^{N_a}. \qquad (16)$$

It is trivial to show from (16) that the expected number of products produced by a country is a monotonically increasing function of the number of capabilities it has[7].

$$\frac{d\overline{k_{c,0}}}{dk_{c,0}^a} = qN_p \left( q \frac{k_{c,0}^a}{N_a} + 1 - q \right)^{N_a - 1} \geq 0. \qquad (17)$$

Next, we calculate the expected ubiquity of a product requiring $k_{p,0}^a$ capabilities. In a model independent form, this can be expressed as the sum of the number of countries with a given number of capabilities times the probability that each one of this exports a product:

$$\overline{k_{p,0}} = \sum_{x=0}^{N_a} \pi \left( c(k_{c,0}^a = x) \rightarrow p(k_{p,0}^a) \right) N_c(k_{c,0}^a = x), \qquad (18)$$

here $N_c(k_{c,0}^a = x)$ is the number of countries that have $x$ capabilities and $\pi \left( c(k_{c,0}^a = x) \rightarrow \right.$ $pkp,0a$ is the probability that a country with $x$ capabilities makes a product that requires $kp,0a$ capabilities. We take (18) into the case where the capabilities that countries have and products require are random by considering $k_{p,0}^a$ as fixed for the product under study, and that the probability for a randomly chosen country to export a product requiring $k_{p,0}^a$ capabilities is given by the probability that it has each of the $k_{p,0}^a$ capabilities that the product requires.

$$\pi \left( c(k_{c,0}^a = x) \rightarrow p(k_{p,0}^a) \right) = \left( \frac{x}{N_a} \right)^{k_{p,0}^a}. \qquad (19)$$

Since in this implementation, the number of capabilities that countries have is in average equal to $N_a r$ and, the fluctuations from this value are not large, we approximate

$$\pi \left( c(k_{c,0}^a = x) \rightarrow p(k_{p,0}^a) \right) = r^{k_{p,0}^a}. \qquad (20)$$

Finally, we consider that the number of countries with $x$ capabilities is given by $N_c$ times a binomial distribution $N(k_{c,0}^a = x) \sim N_c \mathrm{B}(N_a, r)$ and that $r^{k_{p,0}^a}$ comes out of the sum in (18) as it does not depend on the summand $x$. Hence, in this particular case, eqn. (18) simplifies to:

$$\overline{k_{p,0}} = N_c r^{k_{p,0}^a}. \qquad (21)$$

From (21) it is trivial to show that the ubiquity of a product is a decreasing function of the number of capabilities it requires, since $r$ is by definition <1.

Finally, we calculate the average ubiquity of the products exported by a country in a general form as:

$$\overline{k_{c,1}} = \frac{\sum_x^{N_a} \pi \left( c(k_{c,0}^a) \rightarrow p(k_{p,0}^a = x) \right) N_p(k_{p,0}^a = x) k_{p,0}(k_{p,0}^a = x)}{\sum_x^{N_a} \pi \left( c(k_{c,0}^a) \rightarrow p(k_{p,0}^a = x) \right) N_p(k_{p,0}^a = x)}. \qquad (22)$$

Which using the above mentioned algebra and the binomial theorem simplifies to

[7] Notice that for large $N_a$, equation (16) reduces to the exponential form: $\overline{k_{c,0}^p} = N_p \exp(q(k_{c,0}^a - N_a))$

$$\overline{k_{c,1}} = \frac{N_c\left(rq\frac{k_{c,0}^a}{N_a}+1-q\right)^{N_a}}{N_p\left(q\frac{k_{c,0}^a}{N_a}+1-q\right)^{N_a}}. \tag{23}$$

To obtain $k_{c,1}(\overline{k_{c,0}})$ we invert (16) to obtain:

$$k_{c,0}^a(k_{c,0}) = \frac{N_a}{q}\left(\left(\frac{k_{c,0}}{N_p}\right)^{1/N_a}+q-1\right) \tag{24}$$

which we insert into (23), to obtain after some algebra, an expression for the average ubiquity of a country's products as a function of its diversification. This is our prediction regarding the functional form connecting a country's diversification to the average ubiquity of its products.

$$\overline{k_{c,1}} = \frac{N_pN_c}{\overline{\overline{k_{c,0}}}}\left(r\left(\frac{\overline{k_{c,0}}}{N_p}\right)^{1/N_a}+(1-q)(1-r)\right)^{N_a}. \tag{25}$$

Finally, we differentiate (25) with respect to diversity ($k_{c,0}$) to obtain

$$\frac{d\overline{k_{c,1}}}{d\overline{\overline{k_{c,0}}}} = -\frac{N_pN_c}{\left(\overline{k_{c,0}}\right)^2}(1-q)(1-r)\left(r\left(\frac{\overline{k_{c,0}}}{N_p}\right)^{\frac{1}{N_a}}+(1-q)(1-r)\right)^{N_a-1}, \tag{26}$$

demonstrating that ubiquity is negatively related to diversification as long as $q$<1 and $r$<1, proving that in this model the structure of $\boldsymbol{M}_{cp}$ is such that the ubiquity of a country's products decreases with that country's level of diversification.

Next we use the binomial model to calculate the distribution of country diversification and product ubiquity. We do this by using a mathematical identity that connects two random variables, $x$ and $y$, that are related by the function $x=g(y)$. The identity states that, up to a normalization constant, the distribution followed by the random variable $y$ ($P(y)$), will be related to the distribution followed by the random variable $x$ ($f(x)$) by:

$$P(y)\sim f\big(g(y)\big)\frac{dg(y)}{dy}. \tag{27}$$

To calculate the diversification distribution that emerges from the binomial model, we use (24) to find the rate of change of $k_{c,0}^a$ as a function of $k_{c,0}$.

$$\frac{dk_{c,0}^a}{dk_{c,0}} = \frac{1}{qN_p}\left(\frac{k_{c,0}}{N_p}\right)^{(1-N_a)/N_a} \tag{28}$$

Using (27), (28), and the fact that the number of capabilities in a country follows a binomial distribution, we can show that the distribution of diversifications that emerges from the binomial model is given by

$$P(u) = A\frac{N_p}{qu}\binom{N_a}{\frac{N_a}{q}\left(u^{1/N_a}+q-1\right)}r^{\frac{N_a}{q}\left(u^{1/N_a}+q-1\right)}(1-r)^{N_a-\frac{N_a}{q}\left(u^{1/N_a}+q-1\right)} \tag{29}$$

where ($u = k_{c,0}/N_p$) is the fraction of all products exported by a country and $A$ is a normalization factor. Similarly, we can show that in this model the distribution of the fraction of countries that export a product ($v = k^c_{p,0}/N_c$) is given by:

$$P(v) = A\frac{N_c}{v\log(r)}\binom{N_a}{\log(v)/\log(r)}q^{\log(v)/\log(r)}(1-q)^{N_a-\log(v)/\log(r)} \quad (30)$$

This concludes our derivation of the distributions of diversification and ubiquity, up to a normalization factor (eqns. (29) and (30)).

Finally, the predictions of the model regarding the distribution of proximities will be calculated numerically and explored in the next section, where we will bring the model to the data.

## IV. Calibration: Comparing the Model to the Data

In this section, we compare the predictions of the binomial model with the stylized facts laid out in the introduction. The model presented in the previous section has three free parameters, $r$, $q$ and $N_a$ and fitting it to the data requires us to estimate the set of these parameters that provides the best fit.

The first criterion that we use to bring the model to the data is to match the density, or fill, of the network to that observed in the data. The density of a network ($\eta$) is the ratio between the number of links in the network (i.e. the number of 1's in the matrix) and the total number of possible links ($N_c \times N_p$). For the two datasets and thresholds considered, we find that the network densities range between 20% and 8% (see Table 1).

| | SITC4 ($N_c$=129,$N_p$=772) | HS6 ($N_c$=232,$N_p$=5109) |
|---|---|---|
| $\eta$ (R*=1) | 13.53% | 8.54% |
| $\eta$ (R*=0.5) | 19.62% | 12.57% |

Table 1 Network density for the three datasets at $R^*$=1 and $R^*$=0.5.

Since all countries are *ex ante* identical in the binomial model, the density ($\eta$) of the $\boldsymbol{M}_{cp}$ matrix can be calculated as the average fraction of products made by a country. This is equal to the probability that a country will export a good requiring the average number of capabilities that goods require ($qN_a$):

$$\eta = r^{qN_a}. \quad (31)$$

Equation (31) defines a constraint between $r$, $q$, and $N_a$ ensuring that the number of links, the average diversification and the average ubiquity of the networks in the model and the data are the same. Also, this constraint reduces the number of free parameters in the model to two. We choose these parameters as $r$ and $N_a$ and use (31) to solve for $q$ as a function of them:

$$q = \frac{1}{N_a}\frac{\ln(\eta)}{\ln(r)}. \quad (32)$$

Next, we search for values of $r$ and $N_a$ that best reproduce the empirical relationship we described as stylized fact 1, the negative relationship between the average diversification of a country and the average ubiquity of its products (the $k_{c,0}$-$k_{c,1}$ diagram). We do this by using equation (32) to substitute for $q$ in equation (25), and fit the predicted functional form to the empirical data. Figure 6 present these fits and show where in the phase space defined by $r$ and $N_a$ the best fits are found. Fits were determined using least squares, for our two datasets and two cut-offs ($R^*$=0.5 and 1.0).

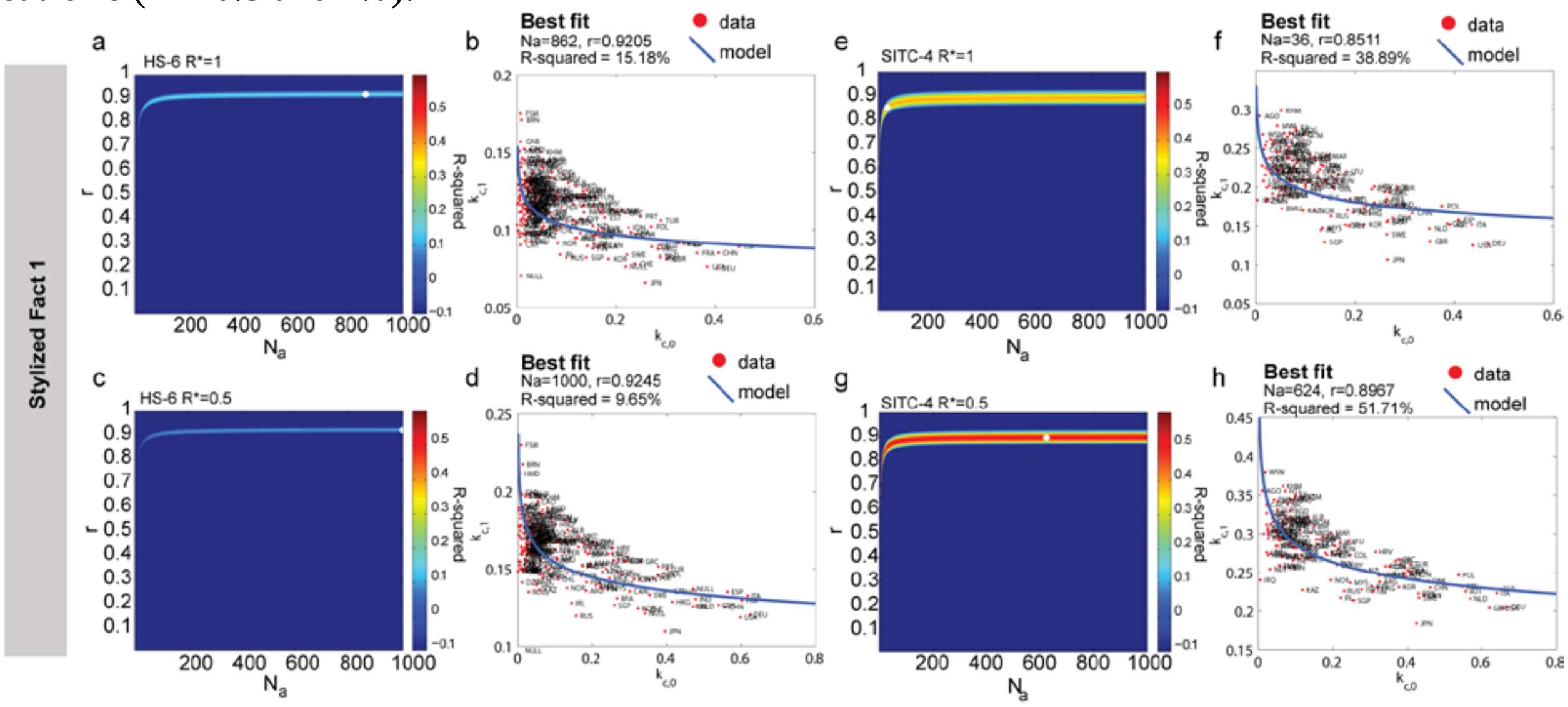

**Figure 6** Calibration of the Binomial Model to the $k_{c,0}$-$k_{c,1}$ diagram. **a** Values of the $R$-squared statistic of the Binomial Model fit to the $k_{c,0}$-$k_{c,1}$ for $r$~[0,1] and $N_a$~[0,1] for the HS-6 dataset and $R^*$=1. **b** Illustration of the best fit of the binomial model to the HS-6 dataset with $R^*$=1. **c** and **d** Same as **a** and **b** but for $R^*$=0.5. **e** and **f** Same as **a** and **b** but for SITC4 data and $R^*$=1. **g** and **h** Same as **a** and **b** but for SITC4 data and $R^*$=0.5.

Figure 6 compares the functional form predicted by the model (solid lines) and the empirical data (points) showing that the model fits the data for a range of $r$ and $N_a$ values. Interestingly, the set of $N_a$ and $r$ values where the best fits are found is similar for all datasets and thresholds and suggests that the model tends to fit the world when the number of capabilities in countries is relatively high ($r$ >0.7) and when the number of capabilities in the world is also relatively large ($N_a$ >30).

Figure 6 shows that the functional form predicted by the binomial model can account for the first of the stylized facts presented in this paper: the negative relationship between diversification and ubiquity present in the data. Yet, since good fits are found for a range of values of $r$ and $N_a$ this first stylized fact can only narrow down the values that these variables can take, but does not provide a complete calibration.

In order to provide a unique set of parameters we consider a second criterion which is based on our fourth stylized fact: The distribution of proximities. We compare the model with the data by searching for the combination of $r$ and $N_a$ values that best reproduces the distribution of proximities observed in the data. We do this by implementing the model numerically and comparing the empirical data and the model using a weighted Kolmogorov-Smirnov Goodness of Fit Test.

Figure 7 summarizes the result of this exercise by showing – for each one of the datasets and two thresholds – the summary of the Kolmogorov-Smirnov test and a comparison between the empirical and theoretical proximity distributions functions, both in terms of frequency and cumulative distributions. First, we observe that for all datasets and threshold there is a region, rather than a point in the parameter space in which the model accurately reproduces the empirically observed proximity distributions. The important point here, however, is that the band of values for which the model can reproduce the proximity distribution partially overlaps the band where the best fits for the $k_{c,0}$-$k_{c,1}$ diagram are found. The intersection between both of these regions defines a narrow range of parameters for which the model most accurately describes the empirical data, suggesting that the empirical data is best fitted using 65 to 80 capabilities and $r$ values between 0.8 and 0.9 (Figure 7 and Table 2).

We consider as significant the fact that the model adjusts the data only for high double-digit values for the number of capabilities because much of our modeling strategies have assumed that the number of factors of production is small. The structure of world exports, however, suggests that it is very difficult to make sense of the empirically observed patterns unless we assume a much larger number of inputs into the production function.

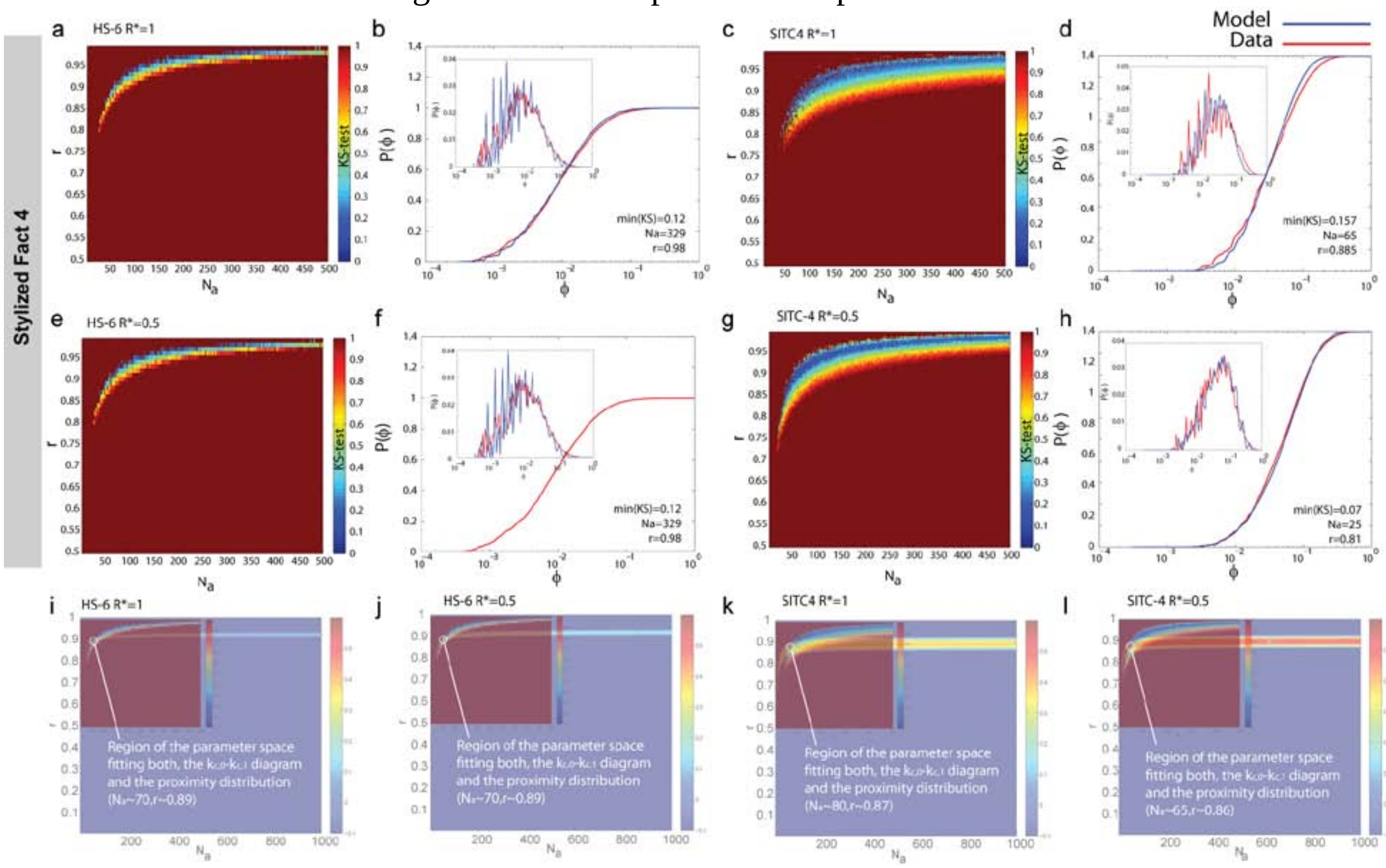

**Figure 7** Calibration of the Binomial Model to the Proximity Distribution. **a** Heat map showing the KS test values obtained after comparing the empirical proximity distributions with that implied by the binomial model (lower values indicate better fits) for HS6 data and R*=1. **b** Cumulative and density probability functions comparing theory and data for the HS6 dataset and R*=1. **c** Same as a but using the SITC-4 dataset and R*=1. **d** Same as b but using the SITC-4 dataset and R*=1. **e** Same as a but using the HS-6 dataset and R*=0.5. **f** Same as b but using the HS-6 dataset and R*=0.5. **g** Same as a but using the SITC-4 dataset and R*=0.5. **h** Same as b but using the SITC-4 dataset and R*=0.5. **i** Overlap between both calibration procedures showing the regions of the $N_a$-$r$ parameter space where the model approximates both, the $k_{c,0}$-$k_{c,1}$ diagram and the proximity distributions that are observed in the data for the HS-6 dataset at R*=1. **j** Same as i but with R*=0.5. **k** Same as i but with the SITC-4 dataset and R*=1. **l** Same as k but with R*=0.5.

Finally, we use this calibration to compare the empirical data with the predictions that the binomial model makes regarding stylized facts 2, 3 and 4, the distributions of diversity, ubiquity and co-export proximity. Here, we take the model into the data in two different ways. First, we implement the model numerically and fit the distributions that emerge from it to a normal, log-normal and Weibull distributions. Next, we use the values of $r$ and $N_a$ that we determined in the calibration procedure to compare the distributions implied by the binomial model (equations (29) and (30)) to the empirically observed distribution. Since the distribution of co-export proximity was used to calibrate the model to the data, this last test is only applied to the ubiquity and diversification distribution.

Figure 7 shows the comparison between the distributions emerging from numerical implementations of the model and a normal, log-normal and Weibull distributions. Here we find the same qualitative behavior. In all cases the normal distribution is the less likely fit and, for the ubiquity and diversification distribution, both the log-normal and Weibull distribution represent better approximations. In the case of co-export proximity, the Weibull distribution always represents the best fit.

| DATASET | $N_a$ | r | q (from (32)) | Kolmogorov-Smirnov Ubiquity distribution | Kolmogorov-Smirnov Diversification distribution |
|---|---|---|---|---|---|
| **SITC-4 R*=0.5** | 65 | 0.86 | 0.1661 | 0.0849 | 0.3962 |
| **SITC-4 R*=1** | 80 | 0.87 | 0.1795 | 0.1189 | 0.3204 |
| **HS-6 R*=0.5** | 70 | 0.89 | 0.2542 | 0.1035 | 0.4025 |
| **HS-6 R*=1** | 70 | 0.89 | 0.3016 | 0.1084 | 0.3368 |

**Table 2 Summary of Calibration**

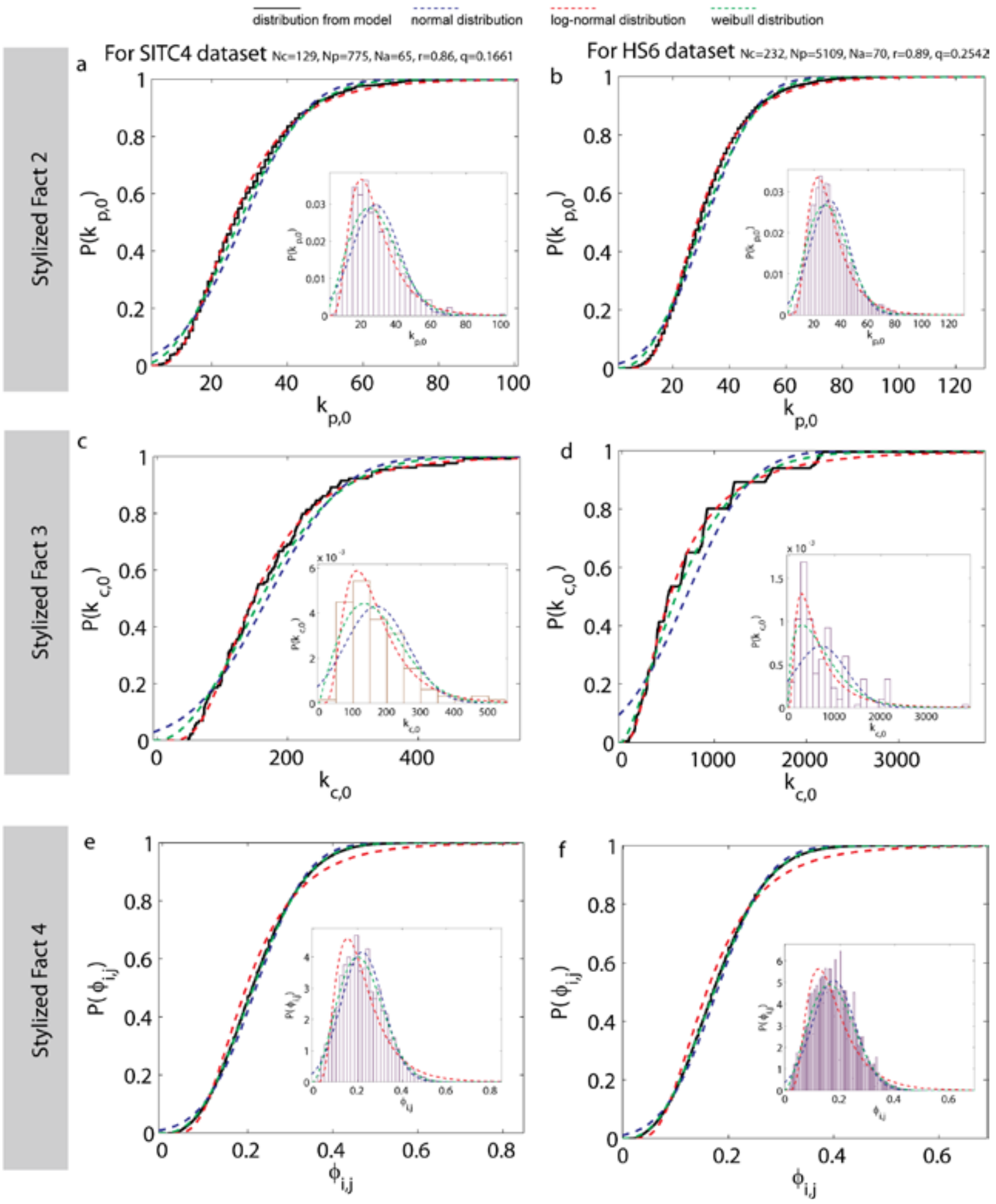

**Figure 8**. Comparison between a numerical implementation of the binomial model and a normal, log normal and Weibull distribution for R*=0.5. **a** Diversification distribution emerging from the parameters estimated for the SITC-4 dataset **b** same as a but for the HS-6 dataset. **c** Ubiquity distribution emerging for the parameters estimated

using the SITC-4 dataset. **d** Same as c but from the HS6 dataset. **e** Co-export proximity distribution emerging from the model and the parameters estimated for the SITC-4 dataset. **f** same a e but for the HS-6 dataset.

Figure 9 compares the empirical ubiquity distribution (probability that a product will be exported by *x* countries) with the prediction from the binomial model (eqn. (30)), showing that the binomial model approximates well the ubiquity distributions observed in the world. Figure 9 also compares the empirical diversification distribution (probability that a country will export *x* products) with the predictions of the binomial model (eqn. (29)). In the case of the diversification distribution, however, the model does not provide an accurate fit. This is because the model implies a range of variation in the diversification of countries that is narrower than that implied in the data. We find this to be a result of the assumption that all countries have the same probability of having a given capability.

We relax this assumption by allowing each country to have its own *r*. We do this by using eqn. (24) to estimate the number of capabilities that a country is expected to have given its diversification and the parameters found in the previous calibration. We interpret this value as $r_c$, the probability that country *c* has a capability, and use this value to reconstruct numerically $\boldsymbol{C}_{ca}$ and implement the model. We compare the distribution of diversification found in the data and the average distribution that comes out of the model after 1,000 numerical simulations and find that these become much closer after incorporating heterogeneity in the capability endowment of countries. Clearly, fitting the distribution of diversification by assuming that each country has a different endowment of capabilities introduces a large number of additional parameters, making our ability to fit much higher. But from an economic point of view, the exercise shows that the high heterogeneity observed in the distribution of diversification can be interpreted as heterogeneity in the distribution of capabilities. This non-normal distribution may be the reflection of non-linearities in the accumulation of capabilities, as we discuss in the next section.

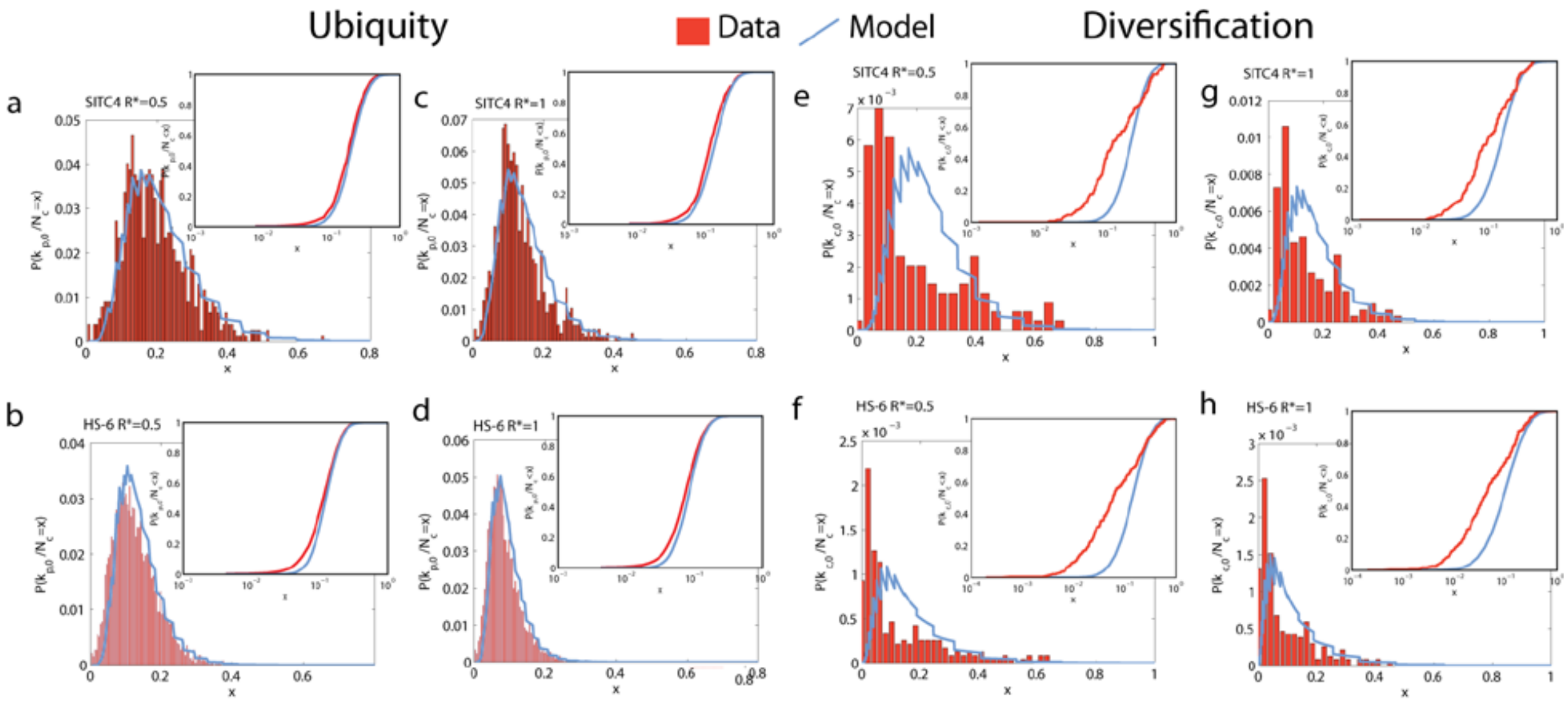

**Figure 9** Comparison between the empirically observed ubiquity and diversification distributions and the one emerging from the binomial model for the parameters found by calibrating the model using the $k_{c,0}$-$k_{c,1}$ diagram and the proximity distribution. For information on the dataset and threshold used in each figure panel see figure. Inset show cumulative distributions **a-d**. Ubiquity distribution **e-h** Diversification distribution.

## V. Implications of the Model

So far, we have introduced a set of four stylized facts about the world's diversity, introduced a general framework that can be used to make predictions regarding the structure of the network connecting countries to the products that they export, shown how to solve it for a particular case and compared its agreements and disagreements with the empirical data. In this section we interpret the framework introduced and relate its structure with that of our world.

### The Quiescence Trap

Our first stop is a purely theoretical prediction about the world that is implied by the model. Equation (16) shows the prediction that the binomial model makes regarding the number of products that a country produces and the number of capabilities that it has. From (16) it is trivial to show that $d^2k_{c,0}/d{k^a_{c,0}}^2 > 0$ for $N_a$>1, indicating that the number of new products that a country can make increases with the number of capabilities in a convex or upward concave form. More to the point, the model predicts increasing returns in product diversity to the accumulation of capabilities for $N_a$>1.

This convexity is valid for all Leontief type operators as long as we assume that there is no correlation between the structure of $\boldsymbol{C}_{ca}$ and $\boldsymbol{P}_{pa}$. Here, we show that this is a more general result by taking the model independent form for $k_{c,0}(k^a_{c,0})$ (eqn. (12)) and noticing that the number of products that require a given number of capabilities $N_p\left(k^a_{p,0}=x\right)$ is by definition independent of the number of capabilities present in a country $\left(k^a_{c,0}\right)$. Moreover, we notice that because of the structure of the Leontief operator, the probability that a country with a given number of capabilities will produce a product requiring x capabilities is given by $\pi\left(c\left(k^a_{c,0}\right) \rightarrow p\left(k^a_{p,0}=x\right)\right)=\left(\frac{k^a_{c,0}}{N_a}\right)^x$ regardless of the structure of the matrices $\boldsymbol{C}_{ca}$ and $\boldsymbol{P}_{pa}$, as long as these are uncorrelated. After these two considerations, we differentiate eqn. (12) with respect to $k^a_{c,0}$ to show that the diversification of a country increases convexly with the number of capabilities it has. The first two derivatives between the diversification of a country and the number of capabilities present in it are:

$$\frac{dk_{c,0}}{dk^a_{c,0}} = \sum_{x=0}^{N_a} x\left(\frac{k^a_{c,0}}{N_a}\right)^{x-1} N_p(k^a_{p,0}=x) \tag{33}$$

$$\frac{d^2k_{c,o}}{d{k^a_{c,0}}^2} = \sum_{x=0}^{N_a} x(x-1)\left(\frac{k^a_{c,0}}{N_a}\right)^{x-2} N_p\left(k^a_{p,0}=x\right). \tag{34}$$

Eqn. (34) shows that the second derivative of diversification with respect to the number of capabilities available in a country is always positive, proving that the convex increase in diversification that is associated with the accumulation of complementary capabilities is a result that does not depend on any assumption regarding which countries (products) have (require) what capabilities.

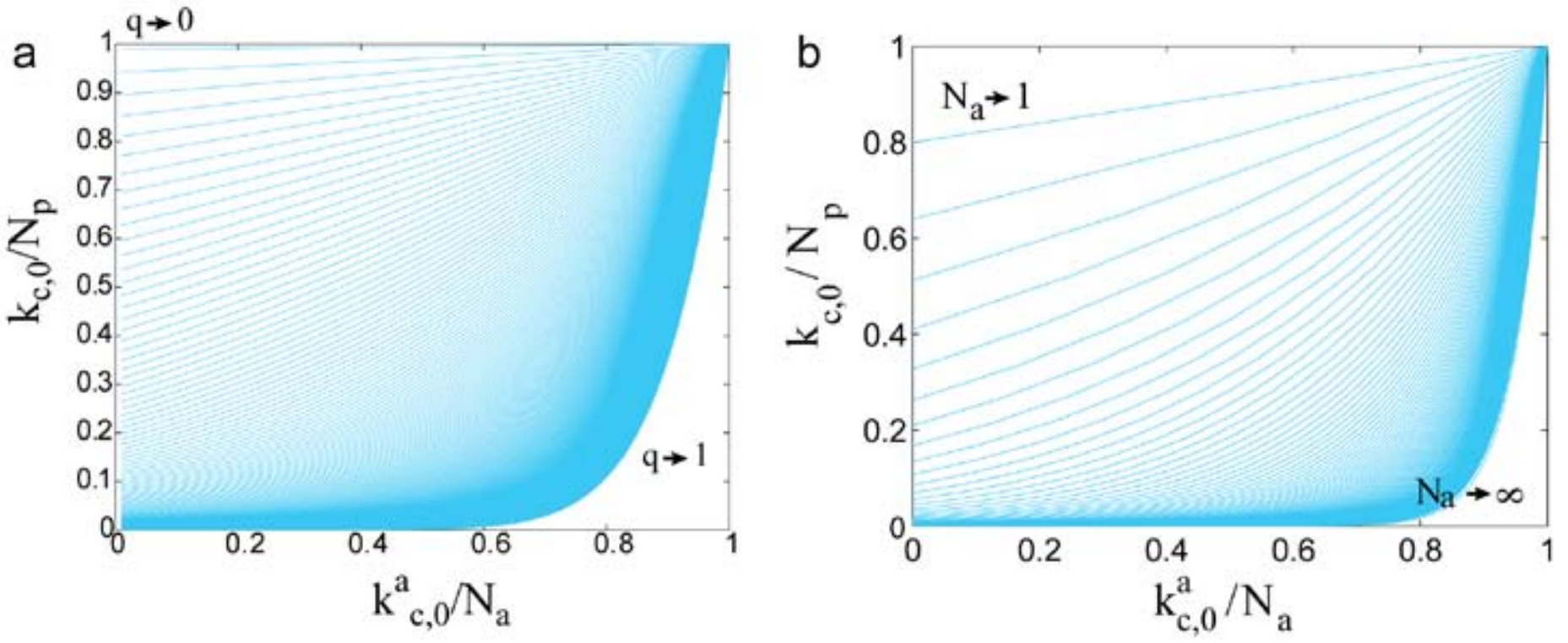

**Figure 10** analytically calculated predictions from the Binomial Model. **a** Fraction of all products that countries can make as a function of the fraction of all capabilities that countries have for values of $q$ ranging from 0 to 1 and $N_a$=50. **b** Fraction of all products that countries can make as a function of the fraction of all capabilities that countries have for $q$=0.2 and $N_a$ ranging from 1 to 1000.

Figure 10 illustrates the implications of this convexity by showing the relationship between the diversity of products and the diversity of capabilities derived from the particular case of the model solved in this paper (eqn. (16)). Figure 10a does this for $N_a$=50 and a range of $q$ values. It is clear from here that the curvature of eqn. (16) becomes more pronounced as $q$ approaches *1*. This curvature has important implications, since the slope of the curve connecting the diversity of outputs to that of inputs represents the number of new products that will become accessible for a country after increasing the number of capabilities it has by a small amount. The model predicts that, as the fraction of capabilities required by the average product increases ($q$ -> 1), the number of new products that become accessible after accumulating a few capabilities becomes small to negligible for countries with only a few capabilities, while at the same time it becomes extremely large for countries with many of them. This implies that in worlds in which products require a large fraction of the total number of capabilities that exist ($q \to 1$), catching up becomes more difficult, since the large fraction of capabilities required by products causes simultaneously, negligible returns for countries with few capabilities and large returns for countries with many of them.

To understand this better consider a world in which products require on average 30 capabilities out of 50 ($q$=3/5). Countries with only five capabilities get no returns for the accumulation of one or two extra capabilities and would likely get no benefit from the accumulation of 10 or even 15 capabilities, since there is no guarantee that the capabilities that they accumulate are exactly those required by the simplest products. In the same world, however, countries with 40 or more capabilities will have large returns to the accumulation of any additional one, since it will be possible to put any new capability that comes along into use in combination with the capabilities already present in that country.

Hence, the model predicts the existence of a quiescence trap, or a trap of development stasis, in which countries with a low diversity of capabilities get stuck. At the same time it predicts that the relationship between the proportion of capabilities the country has and the number of products it is able to make is convex. This convexity increases with both $q$ and $N_a$. Figure 9b shows equation 20 for $q=0.2$ and $N_a$ values in the range [1 1000]. This figure shows that as the number of capabilities in the world increases, the returns to the accumulation of new capabilities become more convex showing that the quiescence trap explained above can also emerge as a consequence of the existence of a large number of capabilities that are highly specific and not only as a result of products requiring a large fraction of the total number of capabilities available.

Finally, we ask the question whether the calibration of the world presented in the previous section implies that our world is one in which the quiescence trap is large or small. Figure 11 shows the relationship between the fraction of capabilities that a country has and the fraction of products that it can make that comes from the calibration performed in the previous section. The calibration suggests that our world is one in which the quiescence trap is strong.

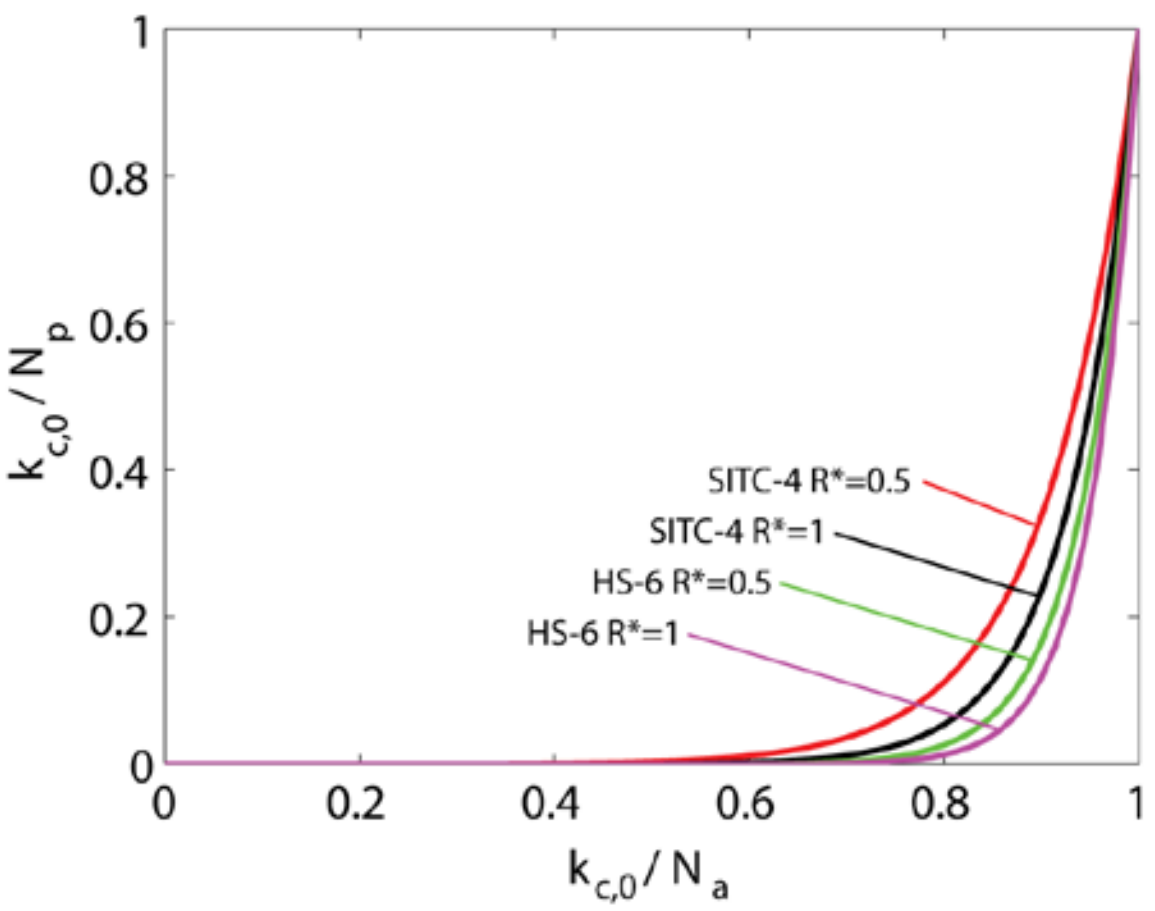

**Figure 11** Relationship between fraction of capabilities that countries have (x-axis) and the fraction of products that countries make (y-axis) for the calibrated model values.

## Concluding remarks

In this paper, we have studied the characteristics of the relationship between products and the countries that make them and presented four stylized facts describing the nature of this structure. First, we find that countries differ not just in how diversified they are, but also in the ubiquity of the products that they export. Moreover, we show that there is a systematic relationship between these two concepts that cannot be explained by the distribution of diversification of countries and the distribution of the ubiquity of products, but speaks to a more fundamental link between the two. Second, we found that the distribution of country diversification, product ubiquities and co-exports is not normally distributed, but that they follow a distribution that can be approximated as a Weibull or log normal distribution.

We propose an analytical framework to account for these stylized facts. We assumed that each product requires a varied and potentially large set of different complementary non-tradable inputs, which we call capabilities. Countries differ in the capabilities that are present in their territory while products differ in the capabilities they require. As a consequence, countries with more capabilities will be more diversified, and products that require more capabilities will be accessible to fewer countries, and hence will be less ubiquitous. Also, countries with more capabilities will be able to make products that require more capabilities, but these are less ubiquitous. This logic explains the negative relationship between the diversification of countries and the average ubiquity of the products that they make.

The theory presented above predicts traps in the process of economic diversification. Countries with few capabilities will be able to make few products and will have scant benefits from accumulating any individual additional capability. This is because the likelihood that a new capability will be able to synergize with existing capabilities and become useful for the production of a new product is low in the absence of the other requisite capabilities. Therefore, the demand for any randomly selected additional capability is likely to be zero in countries with few capabilities. By contrast, countries with many capabilities would be able to produce many new products by combining any new capability with different subsets of the capabilities they already possess. In other words, the model generates increasing returns in terms of diversification to the accumulation of capabilities. Moreover, these increasing returns become more acute as products become more intricate (i.e., as products require a larger proportion of possible capabilities) or when the total number of capabilities in the world is large. Under these conditions, we would not expect the number of capabilities that countries actually have in the world to be a normal distribution, as countries with a larger initial number of capabilities would have found it more advantageous to accumulate more capabilities while those with few capabilities would be trapped in quiescence. So, our finding that we can only explain the distribution of diversification by allowing more diversity in the capability endowment of countries makes sense.

When we calibrate the model to our two different datasets, we find that it is hard to make sense of the observed features of the $\boldsymbol{M}_{cp}$ matrix unless we assume that the number of potential capabilities is in the double digits: between 65 and 80. If our interpretation of the features of the $\boldsymbol{M}_{cp}$ matrix is correct, countries differ not just in the quantity of each capability, but in the variety of capabilities they have. Seen from this perspective, the challenge of development involves solving the coordination problem between the accumulation of additional capabilities and the demand for those capabilities, which presumes the presence of all the complementary capabilities that would be required by a new activity.

The results presented in Hidalgo et al (2007) and in Hausmann and Klinger (2006) show that countries patterns of comparative advantage evolve by moving from existing goods to “nearby” or related goods in The Product Space. This suggests that proximity is related to the similarity of the requisite vector of inputs and that production evolves by minimizing the coordination problem. However, the ability to add a product to the production set of a country depends not only on how close a given product is to an already existing one, but also on how many other capabilities are present in the country and used in other, potentially more distant, products.

The description of the development process that emanates from this paper suggests an important distinction between several dynamics. At the global level, new products and capabilities are created and new ways of making old products are found. In addition, capabilities may become more tradable allowing countries to import inputs that were hitherto non-tradable. At the country level, diversification may increase because entrepreneurs find valuable new combinations of already existing capabilities. Alternatively, new capabilities are accumulated and entrepreneurs search the possible new combinations that the recently added capabilities open up. In addition, in the attempt to make a new product, entrepreneurs identify missing capabilities and act to address the missing input. Finally, the new tradability of a particular input may relax the constraint that had been restricting the development of some products.

Two forces are probably at stake. On the one hand, as products become more complex in terms of the capabilities they require, they become less accessible from the point of view of local production. But as new capabilities become more tradable, manufacturing complexity can be addressed through the international division of the value chain. While it may have been the case at one point in time that to get into shirt manufacturing you needed to master product design, cloth selection and procurement, cutting, sewing, branding, marketing and distribution, now many countries can get into the business by just cutting and sewing to order with additional capabilities added gradually over time.

The quiescence trap generates a convexity of a different in kind from that described by Rosenstein-Rodan (1943) and formalized by Murphy, Schleifer, and Vishny (1989). In those models, the question is one of complementarities among few industries that exhibit economies of scale, such as manufacturing and railroads. A big aggregate push would move the economy to the good equilibrium by coordinating the supply and demand for trains. In such world, central planning may be a possible solution.

In contrast, in the world described in this paper, there are dozens of capabilities, an exponentially growing set of possible combinations between them ($2^{Na}$ to be exact), and incompleteness of the capability set. In such a world, the likelihood that a "big push" will succeed, understood as the provision of a given capability, is lower precisely in the places where the development challenge (or the quiescence trap) is largest, since there will likely be many other missing capabilities that go into making any particular product. In another sense, the model helps clarify the ideas in Hirschman (Albert O. Hirschman, 1958) regarding the creation of disequilibria that would promote backward and forward linkages. In our language, a forward linkage involves the provision of a capability that would then promote the development of an additional product. A backward linkage would be the effective demand for a new capability that emerges from the attempt to make a new product that needs it. Here, forward and backward linkages are the paths towards increasing the variety of capabilities and products. However, the quiescence trap means that this dynamic is more challenging the lower the number of initial capabilities.

This paper opens up several areas of further work. First is the question of why the $\boldsymbol{M}_{cp}$ matrix is triangular rather than diagonal. We have skirted the problem by assuming away the question of quantities, prices, and profits and by having countries produce all the goods for which they have the requisite inputs. A possible explanation is that firm heterogeneity is large relative to the differences in factor prices. Rich countries have more productive firms that can survive in the presence of competition from countries with lower factor costs.

Second, is the question of which are the specific capabilities that seem to go into production. Are they personal skills, non-tradable products, government services, emerging social properties such as rules and norms, or complex combinations of other products such as the capacity to sell physical goods over the Internet? Clearly, they seem to be different from variables such as years of schooling or rule of law, as the correlation between our measures of country complexity and these two variables is low (R is 0.13 and 0.22 respectively for 2006). And yet, our measures of country complexity are robust predictors of future growth.

Third, it would be interesting to explore how the complexity of an economy evolves. What does it take to accumulate new capabilities? How in practice does this occur? Are non-tradable sectors instrumental in creating the domestic demand for capabilities that can then be redeployed in the production of tradables? Did import substitution policies facilitate or hinder the accumulation of capabilities? Does the new tendency to globalize production facilitate development and the accumulation of capabilities by reducing the set of capabilities that need to be present in a given place for production to occur? Is the bi-modal nature of the world's income and diversification distributions a consequence of the convex relationship between capabilities and products and the existence of a quiescence trap?

Fourth, if more complex products require more diversified inputs, then more complex industries should locate only in urban settings where many inputs are available, thus affecting the distribution of urban diversification and of the complexity of the industries they can support. Do we observe a similar $M_{cp}$ matrix when looking across localities within a country, as we saw for Chile?

In sum, the approach we have presented describes development as the process of accumulating a larger variety of capabilities and of expressing them in a wider set of products. The possibilities and pitfalls of this process, which requires the abandonment of a purely aggregative approach, present some of the old questions of economic development in a somewhat different hue.